\documentclass[useAMS,usenatbib,twocolumn]{mn2e}
\usepackage{graphicx}
\usepackage[figuresright]{rotating}
\newcommand {\ggo}{\textquotedblleft}
\newcommand {\ggf}{\textquotedblright}
\newcommand {\hi}{\mbox{H{\sc i}}}
\newcommand {\hii}{\mbox{H{\textsc{ii}}}}
\newcommand {\nii}{\mbox{N{\textsc{ii}}}}
\newcommand {\ha} {H$\alpha$\,\,}
\newcommand {\has} {H$\alpha$}
\newcommand {\kms} {\,km\,s$^{-1}$\,}

\newcommand{\Min}{${}^{\prime}$}

\newcommand{\Deg}{${}^{\circ}$}

\newcommand{\Arc}{''}

\newcommand{\BB}{\textit{\texttt{B}\texttt{\has}\texttt{BAR}\,\,}}
\newcommand{\FM}{\texttt{\textsc{FaNTOmM}}}

\title[\BB: Big \ha kinematical sample of BARred spiral galaxies]{\BB:
Big \ha kinematical sample of BARred spiral galaxies - I. Fabry-Perot
 Observations of 21 galaxies}
\author[O. Hernandez et al.]{O. Hernandez$^{1,2,}$\thanks{Visiting Astronomer, Canada--France--Hawaii Telescope,
operated by the National Research Council of Canada, the Centre
National de la Recherche Scientifique de France, and the
University of Hawaii.}, C. Carignan$^{1,\star}$, P. Amram$^{2,\star}$,
L. Chemin$^{1}$ and O. Daigle$^{1,\star}$  \\
$^1$Observatoire du mont M\'egantic, LAE, Universit\'e de Montr\'eal,
C. P. 6128 succ. centre ville, Montr\'eal, Qu\'ebec, Canada H3C 3J7\\
$^2$Observatoire Astronomique de Marseille Provence et
LAM, 2 pl. Le Verrier,
13248 Marseille Cedex 04, France}

\begin{document}

\date{Accepted - 04/15/2005 . Received 01/30/2005 }

%\pagerange{\pageref{firstpage}--\pageref{lastpage}} \pubyear{2005}

\maketitle

\label{firstpage}

\begin{abstract}
The \ha gas kinematics of twenty-one representative barred spiral
galaxies belonging to the \BB sample is presented. The galaxies
were observed with \FM, a Fabry-Perot integral-field spectrometer, on three
different telescopes. The 3D data cubes were processed through a
robust pipeline with the aim of providing the most homogeneous and
accurate dataset possible useful for further analysis. The data
cubes were spatially binned to a constant signal-to-noise ratio,
typically around 7.  Maps of the monochromatic \ha emission line
and of the velocity field were generated and the kinematical
parameters were derived for the whole sample using tilted-ring
models. The photometrical and kinematical parameters (position angle
of the major axis, inclination, systemic velocity and 
kinematical centre) are in relative good
agreement, except maybe for the later-type spirals. 
\end{abstract}

\begin{keywords}
%\begin{center}
Keywords: galaxies: barred - galaxies: spiral - galaxies: kinematics and
dynamics. Methods: observational. Techniques: radial velocities.
%\end{center}
\end{keywords}

\section{Introduction}

The presence of a bar in disc galaxies seems to be a common
feature. Bars have been recognized in galaxies since the
time of \citeauthor{1918PLick..13...45C} (\citeyear{1918PLick..13...45C}) and \citeauthor{1926ApJ....64..321H} (\citeyear{1926ApJ....64..321H}). In the optical, roughly
30\% of spiral galaxies are strongly barred
(\citeauthor{1963ApJS....8...31D} \citeyear{1963ApJS....8...31D})
while another 25\% are weakly barred. Evidences that bars in
spirals are more obvious in the near-infrared (NIR) than in the
visible go back to Hackwell \& Schweizer (1983). Recent surveys in
the NIR have shown that up to 75\% of high surface brightness
galaxies may have a more or less strong bar (e.g. Knapen, Shlosman
\& Peletier 2000; Eskridge et al. 2000).

When present, such bars will introduce non-circular motions
that should be seen in the radial velocity fields.
Since the kinematics of barred spirals
is different from the one of the more or
less axisymmetric discs, it is important to model them properly if
one wants to derive, as accurately as possible, the overall mass
distribution. This distribution is directly derived from
the knowledge of the circular velocities. Since the gas in
axisymmetric galaxies is nearly on circular orbits, and
its random motions are small compared with  the rotation, its kinematics can be used to derive
rotation curves (RC). However, when a galaxy is barred, the gas
response to the non-axisymmetric part of the potential cannot be neglected.
Thus, a rotation curve derived without correcting for
those non-circular motions cannot be said to represent the
circular motions and be used to derive the mass distribution.

The presence of a bar is expected to leave signatures
mainly in the central regions of the RCs
\citeauthor{1981AJ.....86.1825B} \citeyear{1981AJ.....86.1825B}; \citeauthor{1984PhR...114..321A} \citeyear{1984PhR...114..321A}; \citeauthor{1985AA...150..327C} \citeyear{1985AA...150..327C}; \citeauthor{1997MNRAS.292..349S} \citeyear{1997MNRAS.292..349S}; \citeauthor{2002MNRAS.330...35A} \citeyear{2002MNRAS.330...35A})
which is the region where
the free parameters of the mass models are really constrained
(Blais-Ouellette et al. 1999, 2004; Blais-Ouellette, Amram \& Carignan 2001). Indeed, the parameters of mass models are not constrained by the flat part but by the rising
part of RCs.

%\subsection{Signatures of the Bar in the Velocity Fields.}

Bars in galaxies have very different masses, lengths, axial
ratios, colour distributions, gas content, pattern speeds, shapes
and kinematics.  The determination of the fundamental bar parameters
is a delicate task.  Even if the observational constraints are
numerous, their determinations are rarely unambiguous.
Difficulties come from the fact that

\begin{itemize}
   \item the various galactic
components (bar, disc, spiral arms, bulge, rings) are closely
imbricated in disc galaxies: observations as well as models integrate 
the various components and one cannot observe them separately;
    \item 3D shapes in galaxies are observed projected on the plane of the
sky and their deprojection is not unambiguous, particularly for
barred galaxies;
    \item the parameters of bars depend on the luminous to dark matter ratio
distribution, e.g. galaxies having initially the same disc and the
same halo-to-disc mass ratio but different central haloes
concentrations have very different properties (Athanassoula \&
Misiriotis, 2002).
\end{itemize}

Bars observed from 2D radial velocity fields contain a fraction of
hidden information in tracing the total mass distribution (luminous \& dark) and
they may be directly compared to N-body + SPH simulations. Moreover,
the signature of the bar can appear more significantly through the 2D gaseous velocity field (non symmetrical features due to gas chocs, SF regions of high density, i.e. related to the dissipative nature of the gas) than through the stellar one. So, 2D gaseous velocity fields give a
valuable additional observational information  that will help to disentangle the
parameters of the bar.

For instance, the existence of a velocity
gradient along the minor axis, in both the stellar and the gaseous 2D velocity fields, is not a very good criterion to pick out bars, while the angle between the kinematical major
and minor axes and the twists of the isovelocity contours, are
better criteria (\citeauthor{1981AJ.....86.1825B} \citeyear{1981AJ.....86.1825B}).

Theoretical predictions from N-body simulations (e.g. \citeauthor{1983ApJ...275..529K} \citeyear{1983ApJ...275..529K}; \citeauthor{2000A&A...362..435L} \citeyear{2000A&A...362..435L}; \citeauthor{2002MNRAS.330...35A} \citeyear{2002MNRAS.330...35A}) may be summarized as follow.

When the position angle of the bar is:
\begin{itemize}
    \item roughly parallel to the major axis, the isovelocities show a
characteristic concentration towards the central region due to
the fact that particle orbits are elongated along the bar and the
velocity along an orbit is larger at pericentre than at apocentre;
    \item intermediate between the major and the minor axis positions
angles, the velocity field shows the \ggo \textsf{Z}\ggf\ structure
characteristic of barred galaxy velocity fields (see e.g. \citeauthor{1978ApJ...219...31P} \citeyear{1978ApJ...219...31P}, for NGC 5383);
    \item roughly parallel to the minor axis, the
velocity field shows a sizeable area of solid body rotation in the
inner parts.
\end{itemize}

In the case of a dark halo more concentrated in the central
regions of the galaxy, several of these features remain but some
notable differences appear (Athanassoula \& Misiriotis, 2002).

When the position angle of the bar is:
\begin{itemize}
    \item roughly parallel to the
major axis, the isovelocities show a strong pinching in the
innermost region, on or near the bar minor axis;
    \item intermediate between the major and the minor axis positions angles, the
\ggo \textsf{Z}\ggf\ shape of the velocity field is much more pronounced;
    \item  roughly parallel to the minor axis, the innermost solid-body rotation part
does not show strong differences but as we move away from the
kinematical minor axis the isovelocities show a clear wavy
pattern, indicating that the mean velocity is lower at the ends of
the bar than right above or right below it.
\end{itemize}

This study is dedicated to the kinematics and the dynamics of barred galaxies. The aim
will be to derive the most accurate velocity fields possible for a
representative sample of barred galaxies in order to
analyze their kinematics. The \BB sample should provide
the most homogeneous dataset on barred spiral galaxies to date. Once
this database will be available, the following goal will be to model
those galaxies, extract the non-circular component of the velocities
and thus recover the circular motions and derive proper RCs. Only
then will it be possible to model accurately their mass distributions.

This homogeneous study of the kinematics of twenty-one
nearby barred spiral galaxies, based on the two-dimensional (2D)
kinematics of the \ha gas, is presented
in this paper. Section 2 gives an overview of the observational campaign
and presents the global properties of the \BB sample while
Section 3 discusses the data reduction and especially
the adaptive binning that was performed. In Section 4,
the kinematical parameters are derived and
the FP maps are presented in Section 5. The conclusions can
be found in Section 6 and an appendix presents a short observational
description of the galaxies of the \BB sample.

\section{The \BB sample: Observations}

\label{sec:bbsample}

\subsection{The sample}

The twenty-one galaxies of the \BB sample were selected among northern 
($\delta_{\mathrm{J2000}}$ $\geq +$5.0.) nearby barred galaxies
in de Vaucouleurs et al. (1991, hereafter the RC3) with systemic 
velocities $\leq$ 3000\kms.  Their optical sizes D$_{25}$ were
selected to take advantage of the specific field of view (FOV) of the 
three telescopes used (OmM: D$_{25}$ $\geq$ 3.5\Min; CFHT: 2.0\Min~ $\leq$ D$_{25}$ $\leq$ 4.3\Min and OHP: 2.0\Min~ $\leq$ D$_{25}$ 
$\leq$ 6.0\Min).

\begin{figure*}
\begin{center}
\includegraphics[width=6.5cm]{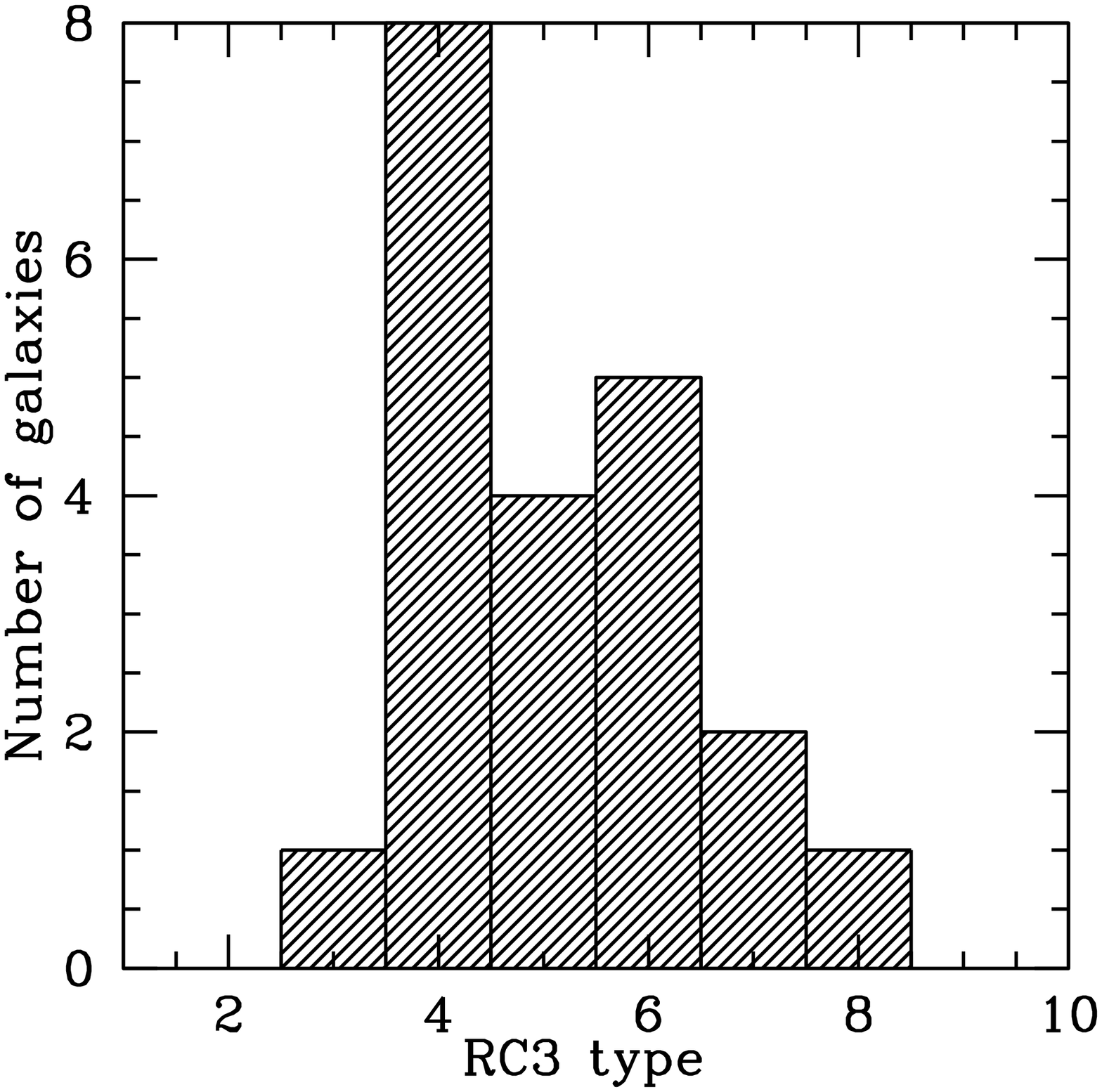} \includegraphics[width=6.5cm]{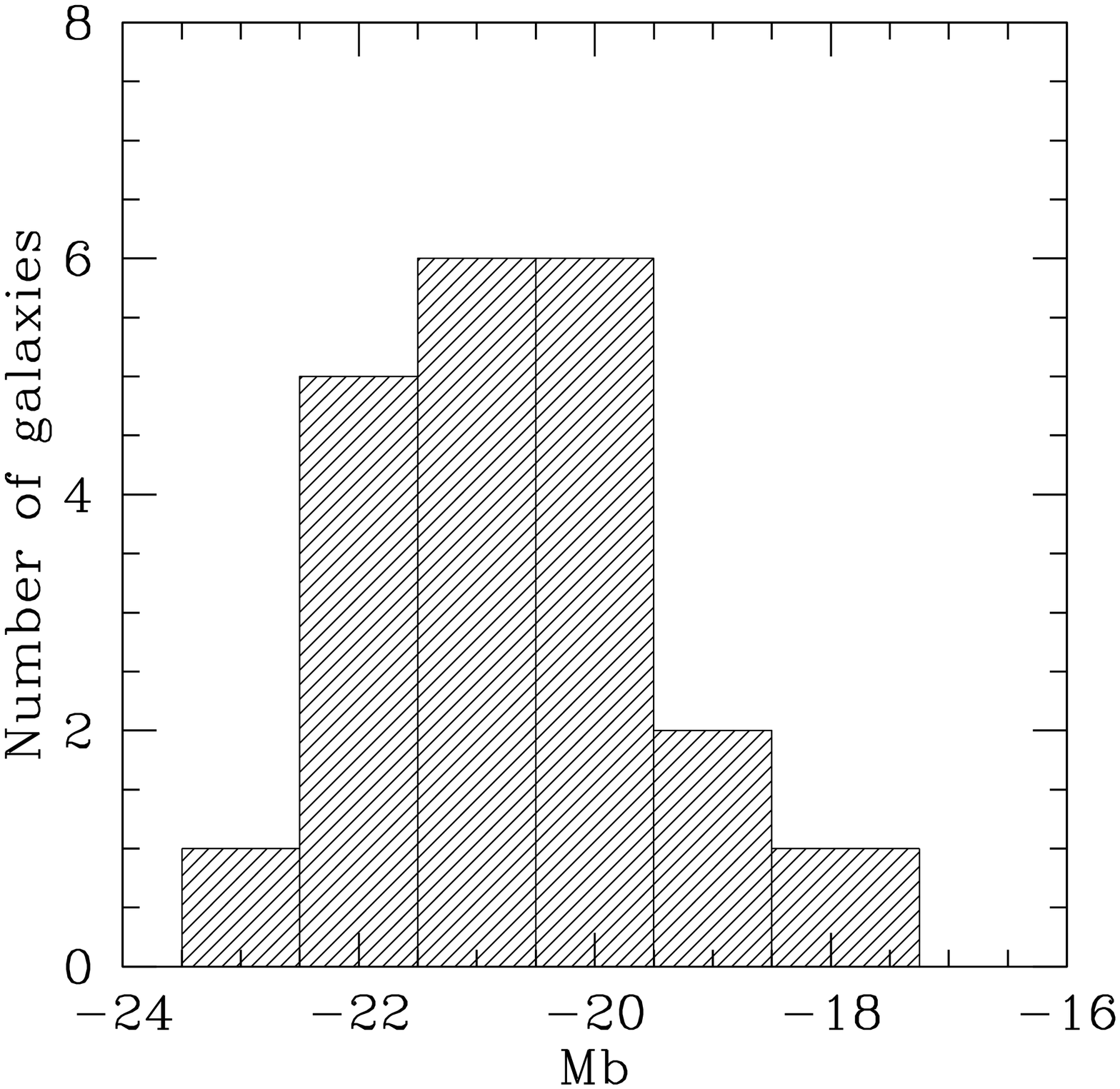}
\end{center}
\caption[Histogram of the sample morphological types and M$_B$]{Histogram of the sample morphological types and M$_B$. \textbf{Left:} The type goes from
SBb (type 3) to SBdm (type 8). \textbf{Right:} M$_B$ varies from -17 to -23.}
\label{histo}
\end{figure*}

To avoid problems related to disc
opacity around H$\alpha$ wavelengths, galaxies with inclination $>$ 
75\Deg~ have been discarded.  The photometrical inclinations $i$ were calculated using
$q=b/a$, the ratio of the minor to the major axis, extracted
from the value of $R_{25}$ given in the RC3 and applying the
following formula:
$$
\cos^2 i = \frac{q^2-q_0^2}{1-q_0^2},
$$ (\citeauthor{1983A&A...118....4B} \citeyear{1983A&A...118....4B})
where $q_0$ is the intrinsic axial ratio of the disc (for an
edge-on system). $q_o$ was not considered as a function of
morphological types since the variation from its nominal value of
$q_0=0.2$ is only significant for late type galaxies (types $\ge$
Sd) which represent only 3 of the 21 galaxies in the sample. %Since
% the inclinations of the galaxies in the \BB sample were chosen
% with i $\leq$ 75\Deg, the problem of extinction along the line of
% sight (as in the case of a nearly edge-on galaxy) and the
% associated uncertainty on the observed velocities are greatly
% reduced. 
The selected galaxies reasonably sample the different morphological Hubble types, from SBb (or SABb) to 
SBdm (or SABdm) and have absolute magnitudes $M_B$ $\leq$ -17. The idea was to try to get a complete and homogeneous coverage of the Hubble sequence. However, due to their poor gas content, only one early type galaxy (SBb) was observed. At the other end of the sequence, only one Sdm was
observed. Figure \ref{histo} gives an histogram of the sample.
 
The galaxies have also been chosen not to be in obvious 
interaction with close companions. To construct mass model and build the gravitational potential in future 
N-body models, galaxies with available \hi\ data and surface
photometry from the literature have been chosen: the photometry comes from Spitzer 3.6 $\mu$m 
and/or J,H,K$_s$ band high resolution images (2MASS) or better
ground images when available (e.g. Knapen et al. 2004 and 2003). Except for two cases 
which are Seyfert galaxies (NGC 6217 and NGC 7479), the galaxies have no nuclear
activity. This confirmation of nuclear activity was not clear at the moment
of the selection of the \BB sample, so it was decided to keep them
in the sample. This will be taken into account when, in further
work, N-body coupled to SPH simulations will be done (forthcoming
papers).

\begin{figure*}
\begin{center}
\includegraphics[width=21.4cm,angle=90]{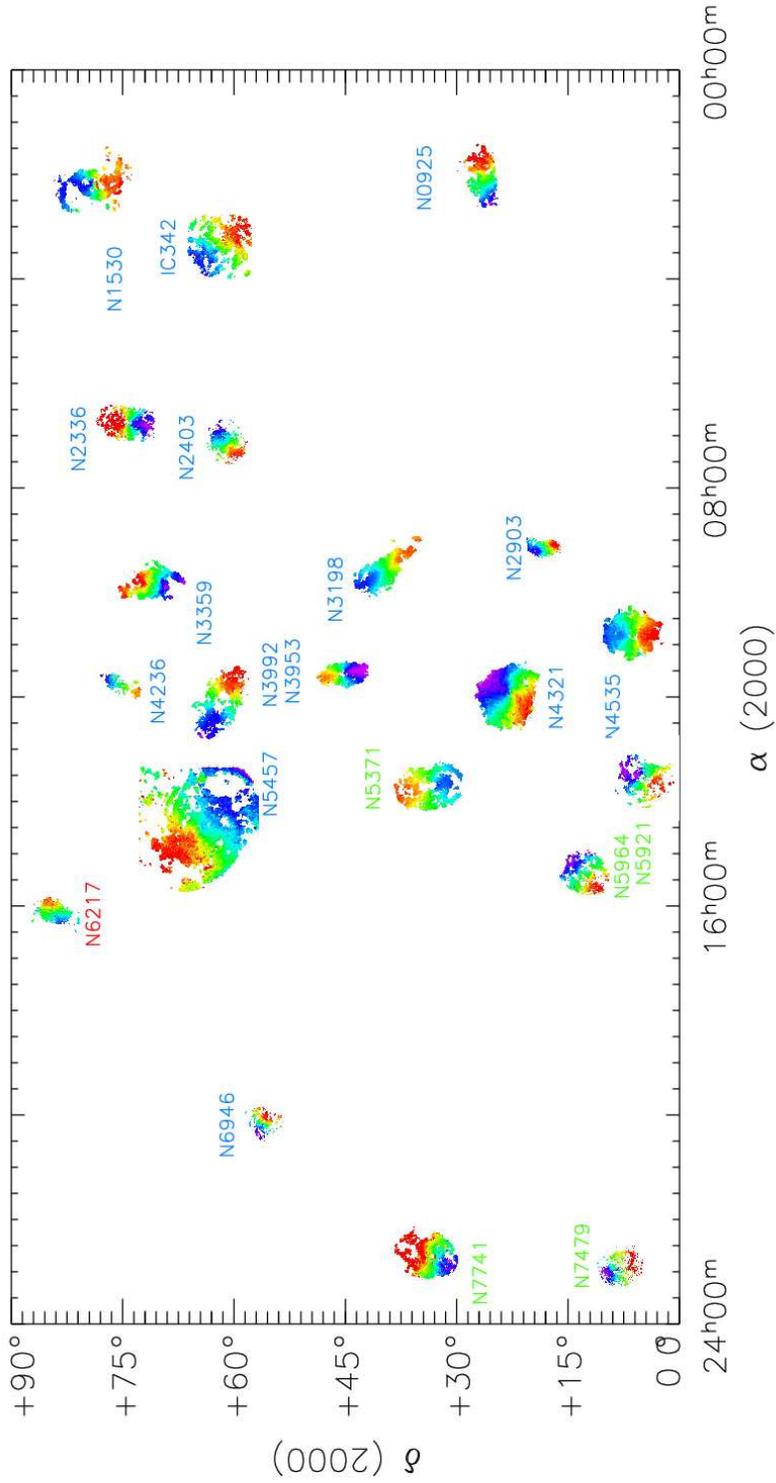}
\end{center}
\caption[The \BB sample sky coverage]{The \BB sample sky coverage. The names in blue correspond
to the observations done at the OmM, in green at the CFHT and in
red at the OHP.  The respective size of the galaxies given in
\emph{kpc} can be directly compared.} 
\label{carte}
\end{figure*}

The global distribution versus morphological type and blue absolute magnitude (M$_B$) are
shown in Fig.~\ref{histo} and the basic information on the objects can be
found in Table \ref{tablesample}. The twenty-one velocity fields of the galaxies are mapped in
Fig.~\ref{carte}.

\begin{table*}
\caption{Observational data for the \BB sample.}
\begin{tabular}{lllllllllllll}
\hline \hline
Galaxy  &   $\alpha$(J2000)&  $\delta$(J2000) &  Type &  D &  $D_{25}^{b,i}$ &  $M_B^{b,i}$ &  $B_T^{b,i}$  &  $L_b^{b,i}$ &  PA$_{\mathrm{b}}$ &  $V_\mathrm{sys}$ & \multicolumn{2}{c}{Image sources}\\
 Name &   hh mm ss &  $^o$ ' " &  RC3 &  Mpc &   $'$ &   & &  \Arc &  \Deg & & B & NIR\\
\hline
 NGC 0925   & 02 27 16.8 &  +33 34 41 & SAB(s)d     &     9.3$^a$ &  11.2 & -20.0 &  10.6 &  56.5 &  112 & 554 & XDSS & Spitzer  \\
 \phantom{NI}IC  0342  &  03 46 49.7 &  +68 05 45 & SAB(rs)cd   &   3.9$^b$  &  27.9 &  -21.6 &  6.4 &  n/a$^c$&  32 &  32 & XDSS & 2Mass \\
 NGC 1530   & 04 23 28.5 &  +75 17 50 & SB(rs)b     &     36.6$^b$ &  4.8 &  -21.3 &  11.5 &  137.0$^c$&  116 &  2460 & XDSS & 2Mass\\
 NGC 2336  & 07 27 04.5 &  +80 10 41 & SAB(r)bc   &    22.9$^b$  &  5.2 &  -22.1 &  11.1 &  72.8 &  5&  2196 & XDSS & 2Mass \\
 NGC 2403   & 07 36 54.5 &  +65 35 58 & SAB(s)cd    &     4.2$^a$   &  21.4 &  -19.7 &  8.5 &  n/a$^c$&   n/a & 132  & XDSS & Spitzer\\
 NCG 2903   & 09 32 09.7 &  +21 30 02 &  SB(s)bc     &    6.3$^b$      &  10.0 &  -19.8 &  9.13 &  143.4 &  26&  554 & XDSS & 2Mass\\
 NGC 3198 &   10 19 54.9 &   +45 33 09 &  SB(rs)c     &    14.5$^a$  &    7.8 &   -20.2 &   11.1 &   81.8 &   12&  660 & XDSS & Spitzer\\
 NCG 3359  &  10 46 37.7 & +63 13 22 &  SB(rs)c     &  19.2$^b$  &  7.2 &  -20.4 &  10.8 &  87.0 &  15& 1013& XDSS & 2Mass\\
 NGC 3953  &  11 53 49.5 &  +52 19 39 &  SB(r)bc     &  17.0$^b$ &  5.2 &  -20.6 &  10.5 &  70.6 &  55&  1054& XDSS & 2Mass\\
 NGC 3992  &   11 57 36.0 &  +53 22 28 &  SB(rs)bc    &  17.0$^b$ &   6.6 &  -20.7 &  10.4&  136.5 &  37&  1051& XDSS & 2Mass\\
 NGC 4236   &  12 16 42.1 &  +69 27 45 &  SB(s)dm     &  2.2$^b$&  16.4 &  -17.3 &   9.5 &   170.7 &   143&  2& XDSS & 2Mass\\
 NGC 4321   &    12 22 55.2 &   +15 49 23 &    SAB(s)bc     &   16.1$^a$  &   7.4 &   -22.1 &   10.0 &   53.4&   107&  1590& Kpn04 & Spitzer \\
 NGC 4535   &  12 34 20.3 &   +08 11 53 &    SAB(s)c     &   16.0$^a$  &   6.9 &   -22.0 &   10.6 &   70.0$^c$ &   45&  1966 & Kpn04 & Kpn03 \\
 NGC 5371   &     13 55 40.6 &   +40 27 44 &    SAB(rs)bc   &   37.8$^b$  &   4.0 &    -21.6 &   11.3 &  47.2 &   97&  2558& XDSS & 2Mass\\
 NGC 5457  &     14 03 12.5 &   +54 20 55 &    SAB(rs)cd   &   7.4$^a$   &   30.2 &    -22.7 &   8.3 &  86.5 &   84&  231& XDSS & 2Mass\\
 NGC 5921   &    15 21 56.4 &   +05 04 11 &    SB(r)bc     &   25.2$^b$  &   4.9 &   -20.7 &   11.3 &  73.5 &   21&  1480& XDSS & 2Mass\\
 NGC 5964   &    15 37 36.3 &   +05 58 28 &   SB(rs)d     &   24.7$^b$  &   4.0 &   -19.5 &   12.5 &  56.0$^c$ &  150&  1447& Kpn04 & Kpn03 \\
 NGC 6217   &    16 32 39.2 &   +78 11 53 &    (R)SB(rs)bc* &    23.9$^b$&   3.6 &   -20.2 &   11.7 &  68.8 &  153&  1359& XDSS & 2Mass\\
 NGC 6946   &    20 34 52.0 &   +60 09 15 &    SAB(rs)cd   &   5.5$^b$    &    14.9 &   -20.8 &   7.92 &   34.5&  166 &  46& Kpn04 & Spitzer \\
 NGC 7479   &    23 04 57.1 &   +12 19 18 &    SB(s)c* &     32.4$^b$ &   3.8  &   -21.1 &   11.4 &   114.7&   2&  2382& XDSS & 2Mass\\
 NGC 7741   &    23 43 54.0 &   +26 04 32 &    SB(s)cd &     12.3$^b$  &    3.8 &   -18.8 &   11.7 &   89.0 &   103&  750& Kpn04 & Kpn03 \\
\hline 
\end{tabular}
\begin{flushleft}
D : distances in Mpc \\
\hspace{0.25cm}$a$ - distances calculated from Cepheids (NGC 0925 - \citeauthor{1994AAS...185.2407S} (\citeyear{1994AAS...185.2407S}), NGC 2403 -
\citeauthor{Freedman88} (\citeyear{Freedman88}), NGC 3198 - \citeauthor{1999ApJ...514..614K} (\citeyear{1999ApJ...514..614K}), NGC 4321 - \citeauthor{1996ApJ...464..568F} (\citeyear{1996ApJ...464..568F}),
NGC 4535 - \citeauthor{1999ApJ...521..155M} (\citeyear{1999ApJ...521..155M}), NGC 5457 - \citeauthor{1996ApJ...463...26K} (\citeyear{1996ApJ...463...26K})).  \\
\hspace{0.25cm}$b$ - distances are based on velocities, an assumed value of the Hubble Constant of 75\kms$Mpc^{-1}$ , and the
model that describes the velocity perturbations in the vicinity of the Virgo Cluster, c.f. \citeauthor{1988JBAA...98..316T} (\citeyear{1988JBAA...98..316T}). \\
$D_{25}^{b,i}$ : optical diameter at the 25 magnitude/arcsecond$^2$ in B.\\
$M_B^{b,i}$ : absolute magnitude in B. \\
$B_T^{b,i}$ : apparent magnitude in B. \\
$L_b^{b,i}$ : bar lengths from Martin 1995, all adjusted for the effects of projection and obscuration. $^c$ values not available in Martin (1995) are calculated from isophote fitting (except for NGC 1530 where the data are from  Zurita et al. \citeyear{2004A&A...413...73Z}).\\
PA$_b$ : PA of the bar. \\
$V_{sys}$ : systemic velocities provided by Tully (1988).\\
% $^b$ - Distances are based on velocities, an assumed value of the Hubble Constant of 75\kms$Mpc^{-1}$ , and the
% model that describes the velocity perturbations in the vicinity of the Virgo Cluster, c.f. \citeauthor{1988JBAA...98..316T} (\citeyear{1988JBAA...98..316T}).\\
The symbol * refers to a nuclear activity. \\
Kpn04 refers to Knapen et al. 2004. Kpn03 refers to Knapen et al. 2003.
\end{flushleft}
\label{tablesample}
\end{table*}

\subsection{Observing runs}

The observations were obtained using \FM\footnote{\FM\ stands for \textbf{Fa}bry-Perot 
of \textbf{N}ew \textbf{T}echonolgy of the \textbf{O}bservatoire 
du \textbf{m}ont \textbf{M}\'egantic (http://www.astro.umontreal.ca/fantomm).} which is a permanent instrument on the OmM telescope, and a visitor instrument on various other telescopes. \FM\, is  a wide integral field
spectrometer made of an image photon counting system (IPCS), a scanning Fabry Perot (FP) and an
interference filter. The photocathode used has a high quantum efficiency
($\sim$ 30\% at \has). \FM\ is coupled to the focal reducers of the
telescopes used (see details in in Table \ref{fmcarac}).

The IPCS has a third generation photocathode with high quantum
efficiency over a large wavelengths range (Hernandez et al. 2003
and \citeauthor{2002PASP..114.1043G} \citeyear{2002PASP..114.1043G}). This camera is very efficient to reach a
good Signal-to-Noise ratio (S/N) for objects with very faint
fluxes since, compared with CCDs, it has no read-out noise. Its
multiplex mode also allows a rapid and efficient suppression of
the OH sky lines since their variations can be averaged out. \FM\
was used in its low spatial resolution mode of 512 $\times$ 512 pixels$^2$
(instead of 1024 pixels$^2$).

\begin{table*}
\caption{\FM\ characteristics on various telescopes.}
\begin{center}
\begin{tabular}{llcccc}
\hline \hline
Telescope & Focal Reducer & $F/D$ & pixel size & FOV & FOV (vign.) \\
Name  &   & & (\Arc) & (\Min) & (\Min) \\
\hline
OmM  & \textsc{panoramix} & 2.3 & 1.61 & 19.4 & 19.4 \\
OHP  & \textsc{cigale} & 3.92 & 0.68 & 8.2 & 5.5 \\
CFHT  & \textsc{mos} & 2.96 & 0.48 & 5.8 & 3.9 \\
\hline
\end{tabular}
\end{center}
\begin{flushleft}
$F/D$ represents the ratio focal
length over telescope diameter. The pixel size after binning is $2 \times 2$, the original GaAs system
providing 1024$\times $1024 px$^2$. FOV is the diagonal Field Of View of the detector.
FOV (vign.) represents the effective unvignetted FOV (whithout the vignetting due to the filter 
used).
\end{flushleft}
\label{fmcarac}
\end{table*}

The observations of the sample were spread over nine different
observing runs over a three years period. Six runs were at the
1.6m of the Observatoire du mont M\'egantic (OmM), one at the
1.93m of the Observatoire de Haute-Provence (OHP) and two at the
the 3.6m of the Canada-France-Hawaii Telescope (CFHT). 
% \FM\ has
% been attached at the Cassegrain focus of the three telescopes, the
% different FOVs on various telescopes are given in Table
% \ref{fmcarac}.
Various FP interferometers were used in order to fit the
adequate spectral resolution. The interference orders vary from
p=765 to p=1162, calculated for $\lambda_0$=6562.78\AA\ (see Table
\ref{JOFP}).

All the calibrations were done using the same neon lamp (see below for
more details on the data reduction). With the rapid analogic
detector mode, calibrations were done in less than a minute. This
allowed to perform as much calibrations as needed during the runs
with very little overhead.
A new bank of interference filters was also used covering a velocity range from 0 to
10,000\kms (from 6562\AA\ to 6785\AA, with $\Delta\lambda\sim15$\AA).

\begin{center}
%\begin{sidewaystable*}
\begin{table*}
%\caption{\, Journal of the Fabry Perot Observations\  
\caption{Journal of the Fabry Perot Observations.}
%\begin{flushleft}
\label{JOFP}
%\begin{minipage}[c][150mm]{\textwidth}
%\centering
\begin{tabular}{c|c|ccc|cc|cccc|cc}
\hline \hline
   Galaxy &   Date & \multicolumn{3}{c}{  Filter}  & 
\multicolumn{2}{c}{  Exposure} & \multicolumn{4}{c}{  Fabry-Perot} & 
\multicolumn{2}{c}{  Sampling} \\
  Name  &   &   $\lambda_c^{(4)}$ &   $\Delta\lambda^{(5)}$ & 
  T$^{(6)}$ &    t$_{tot}^{(7)}$ &   t$_{ch}^{(8)}$ & 
  p$^{(9)}$ &   FSR$^{(10)}$ &   \textbf{\textit{F}}$^{11}$ & 
  \textbf{\textit{R}}$^{(12)}$ &   nch$^{(13)}$ & 
  stp$^{(14)}$ \\ %&   Px$^{(15)}$ \\
\hline
  NCG 0925$^{(1)}$  &   02/11/02 &   6584   &   15 &   75 &   132  &   2.75&   765 &   391.77 &   16 &   12240 &   48 &   0.18  \\
 \phantom{NI}IC  0342$^{(1)}$  &  02/11/03 &  6578 &   15 &  60 &  144  &  3 &  765 &  391.77 &  16 &  12240 &  48 &  0.18  \\
  NGC 1530$^{(1)}$  &   03/01/30 &   6622 &   15 &   70 &   240  &   5 &   899 &   333.36 &   20 &   17980 &   48 &   0.15  \\
  NGC 2336$^{(1)}$  &   02/11/10 &   6617 &   15 &   69 &   240 &   5 &   765 &   391.77 &   16 &   12240 &  48 &   0.18  \\
  NGC 2403$^{(1)}$  &   02/11/17 &   6569 &   10 &   50 &   120  &   3 &   765 &   391.77 &   14 &   10670 &   40 &   0.21  \\
  NCG 2903$^{(1)}$  &   03/02/08 &   6599 &   15 &   74 &   180  &   3.75 &   899 &   333.36 &   20 &   17980 &   48 &   0.15 \\
  NGC 3198$^{(1)}$  &   03/03/06 &   6584 &   15 &   75 &   260  &   5 &   899 &   333.36 &   23 &   20976 &   52 &   0.14 \\
  NCG 3359$^{(1)}$  &   01/11/17 &   6569  &   10 &   50 &   130  &   3.25 &   765 &   391.77 &   14 &   10670 &   40 &   0.21 \\
  NGC 3953$^{(1)}$  &   03/02/26 &   6599 &   15 &   74 &   273  &   5.25 &   899 &   333.36 &   23 &   20976 &   52 &   0.14 \\
  NGC 3992$^{(1)}$  &   03/02/27 &   6599 &   15 &   74 &   260  &   5 &   899 &   333.36 &   23 &   20976  &   52 &   0.14 \\
  NGC 4236$^{(1)}$  &   04/02/27 &   6578 &   15 &   60&   182  &   3.5 &   899 &   333.36 &   23 &   20976  &   52 &   0.14 \\
  NGC 4321$^{(1)}$  &   03/02/25 &   6605 &   15 &   75 &   260  &   5 &   899 &   333.36 &   23 &   20976  &   52 &   0.14 \\
  NGC 4535$^{(1)}$  &   03/03/06 &   6617 &   15 &   69 &   156  &   3 &   899 &   333.36 &   23 &   20976  &   52 &   0.14 \\
  NGC 5371$^{(2)}$  &   03/04/04 &   6622 &   15 &   70 &   88  &   1.83 &   899 &   333.36 &   16 &   17980 &   48&   0.15\\
  NGC 5457$^{(1)}$  &   03/02/28 &   6578  &   15 &   60 &   260  &   3.5 &   899 &   333.36 &   23 &   20976  &   52 &   0.14 \\
  NGC 5921$^{(2)}$  &   03/04/09 &   6599 &   15 &   74 &   120  &   2.5 &   899 &   333.36 &   16 &   17980 &   48&   0.15\\
  NGC 5964$^{(2)}$  &   03/04/08 &   6599 &   15 &   74 &   120  &   2.5 &   899 &   333.36 &   16 &   17980 &   48&   0.15 \\
  NGC 6217$^{(3)}$  &   01/10/17 &   6595 &   10 &   60 &   72 &    3 &   793 &   377.94 &   11 &   8722&   24 &   0.34 \\
  NGC 6946$^{(1)}$  &   02/11/19 &   6569 &    10 &   50 &   120 &    2 &   765 &   391.77 &   14 &   10670 &   40 &   0.21\\
  NGC 7479$^{(2)}$  &   02/10/04 &   6617 &   15 &   69 &   60 &    6 &   1162 &   257.92 &   11 &   12782  &   24 &   0.23 \\
  NGC 7741$^{(2)}$  &   02/10/06 &   6584 &   15 &   75 &   60 &   6 &   1162 &   257.92 &   11 &   12782  &   24 &   0.23 \\
\hline
\end{tabular}\\
\begin{flushleft}
$^{(1)}$ OmM : Observatoire du mont M\'egantic, Qu\'ebec, Canada. 1.6m telescope. \\
$^{(2)}$ CFHT : Canada-France-Hawaii Telescope, Hawaii, USA. 3.6m telescope.\\
$^{(3)}$ OHP : Observatoire de Haute-Provence, France. 1.93m telescope.\\
$^{(4)}$ $\lambda_c$ : filter central wavelength in \AA \\
$^{(5)}$ $\Delta\lambda$ = FWHM : filter Full Width at Half Maximum in \AA \\
$^{(6)}$ T : filter maximum transmission at $\lambda_c$\\
$^{(7)}$ t$_{tot}$ : total exposure time in minutes\\
$^{(8)}$ t$_{ch}$ : exposure time per channel in minutes\\
$^{(9)}$ p : Fabry-Perot interference order at \ha \\
$^{(10)}$ FSR : Fabry-Perot Free Spectral Range (FSR) at \ha in \kms \\
$^{(11)}$ \textbf{\textit{F}} : mean \textit{Finesse} through the field of view\\
$^{(12)}$ \textbf{\textit{R}} : resolution for a signal to noise ratio of 5 at the sample step\\
$^{(13)}$ nch : number of channels done by cycle in multiplex observations \\
$^{(14)}$ stp : wavelength step in \AA \\
\end{flushleft}
%$^{(15)}$ Px = Pixel size after binning 2*2, the original GaAs system providing 1024$\times $1024 px$^2$
  
%\end{flushleft}
%\caption{Journal of Fabry-Perot Observations}
\end{table*}
%\end{sidewaystable*}
\end{center}
\section{Data reduction}

The reduction of the data cubes was performed using the package
ADHOCw (Boulesteix 2004 and \citeauthor{1992A&AS...94..175A} \citeyear{1992A&AS...94..175A}) rewritten with large improvements under the IDL package (Daigle, Carignan \& Hernandez 2005). The major improvements are the following:
\begin{itemize}
    \item the elementary interferograms (elementary images
obtained with an exposure time ranging from 10 to 15 seconds,
depending on the sky transparency conditions and on the number of
scanning steps) were corrected from sky fluctuations before
summation;
    \item adaptive Hanning smoothing was performed in order to increase the S/N over the field;
    \item World Coordinates System (WCS) astrometry on the images was performed;
    \item strong and robust OH night sky lines removal was used;
    \item full automated and reproductible reduction and data analysis were performed.
\end{itemize}

\subsection{Phase Calibration}

Raw interferograms must be corrected to obtain data cubes sorted
in wavelengths. This operation is called the \ggo phase calibration\ggf\ 
or wavelength calibration. Those calibrations are obtained by
scanning the narrow Ne 6599 \AA\ line under the same conditions as
the observations.  Two phase calibrations are done, one before
and the other after the exposure. Using the mean of the
calibrations, a \ggo phase map\ggf\ is computed. It indicates the scanning
step at which is observed the maximum of the interference pattern
inside a given pixel. The FP formula below, giving the
shape of the interference pattern on the detector as a function of
the observed wavelength, allows us to find the observed
Doppler-shifted wavelength $\lambda$ at each point by comparison
with $$p\lambda=2ne\cos\theta$$ where $p$\, is the interference
order at $\lambda_0$ (here 6562.78\AA), $n$ the index of the
medium, $e$ the distance between the two plates of the FP and $\theta$ the incidence angle on the FP
(angular distance on the sky). An uncertainty remains since the
velocity is only known modulo the Free Spectral Range (column 10
in Table \ref{JOFP}, note 11). This ambiguity is easily solved by
using comparisons with long-slit spectroscopy or 21 cm \hi\ line
data to provide the zero point of the velocity scale. 
However, this means that when the redshift emission line of the galaxy is
far from the calibration line, absolute values of the systemic
velocity could be wrong (which is not a problem since we are
mainly interested in relative velocities for our kinematical
studies). In such cases, two ways are in
development: a correction using the dispersion in the multi-layer,
semi reflective high Finesse coating, which is hard to model for
high multilayers coating, or/and an absolute calibration (in
development) done at the scanning wavelength. Nevertheless,
the \textit{relative} velocities with respect to the systemic
velocity are very accurate, with an error of a fraction of a
channel width (${\rm <3 \, kms^{-1}}$) over the whole field. In
this study, systemic velocities of the sample, presented in Table
\ref{tablesample}, were taken directly from Tully (1988).

The signal measured along the scanning sequence was separated in
two parts: (1) an almost constant level produced by the continuum
light in a narrow passband around \ha and (2) a varying part produced
by the \ha line (referred hereafter as the monochromatic map). After
this calibration step, an adaptive binning was performed.

\subsection{Adaptive binning and maps}

A Hanning smoothing was performed on all the data cubes along the
spectral axis. The Hanning smoothing can suppress the problems
connected with the frequency response (artifacts in the spectra
caused by the sampling and the Fourier transform) of the spectra
as a real function of finite length. The strong OH night sky lines
passing through the filter were reconstructed into a cube and
subtracted from the galaxy's spectrum (Daigle, Carignan \& Hernandez 2005).

In order to increase the signal-to-noise ratio (S/N), an adaptive
spatial smoothing, based on the 2D-Voronoi tessellations method
(\citeauthor{2002gtd..conf..515C} \citeyear{2002gtd..conf..515C}) was also applied to the 3D data cubes
(Daigle, Carignan \& Hernandez 2005) before producing the
monochromatic images and velocity fields.  Each pixel was binned
to reach a S/N of typically 5 to 10, depending on the
observation conditions and the morphological type of the galaxy. 
This clever smoothing is effective in low S/N regions.
First, in high S/N regions (S/N value superior to a fixed limit of
5, 7 or 10), the smoothing will not act and a bin is only one
pixel. This will ensure to have the best spatial resolution
possible in high S/N regions. This differs from the classical
gaussian smoothing (\citeauthor{2003A&A...399...51G}
\citeyear{2003A&A...399...51G}, Zurita et al. 2004) where the
kernel used will do the mix between a pixel and the adjacent one,
and will cause a cross-pollution between the two regions. Second,
for low S/N regions, pixels are binned until the S/N required is
reached or the size of the resultant bin is reached (typically
30~$px^2$). This is very useful in interarm regions where the
signal is dominated by the diffuse \ha and not by \hii\ regions.
Thus, velocity maps have the best possible coverage without
loosing any spatial resolution in the high S/N regions.

Finally, the intensity-weighted mean (barycentre) of the \ha
emission line profile was converted into wavelength and then in
heliocentric radial velocity. Monochromatic images
were obtained by integrating the \ha profiles.

% \ref{fmcarac}

\subsection{WCS astrometry}

\textsc{SAOImage ds9} developed by the Smithsonian Astrophysical
Observatory (\citeauthor{1999adass...8..429J} \citeyear{1999adass...8..429J}) has been used to find the correct astrometry
of the monochromatic and \ha images. \textsc{Karma} (\citeauthor{1996adass...5...80G} \citeyear{1996adass...5...80G}) and its
routine \textsc{kpvslice} have been used to apply a co-ordinate system header to
all images and  data cubes. Systematic comparison between Ks band and XDSS Blue
band images and the field stars in rough continuum images (with
no adaptive binning) were made in order to find the correct WCS for each images. For the 21 galaxies of the \BB sample, stars were easily
found in rough continuum images.

\section{Data analysis}

For each galaxy in the sample, Figures \ref{n0925} to \ref{n7741} provide: the blue band image (XDSS or \citeauthor{2004A&A...426.1135K} \citeyear{2004A&A...426.1135K}), the
NIR image (SPITZER 3.6$\mu$m, \citeauthor{2003MNRAS.344..527K} \citeyear{2003MNRAS.344..527K} or 2MASS image), the \ha monochromatic image, the velocity field and the Position-Velocity
(PV) diagram.

Once the astrometry done on all the images and data cubes, the
kinematical parameters were derived using \textsc{Gipsy} and
\textsc{Karma}.

\label{sc:gipsy}
The \textsc{ROTCUR} routine in the \textsc{Gipsy} package was used
to find the kinematical parameters of the galaxies studied.
ROTCUR (\citeauthor{1987PhDT.......199B} \citeyear{1987PhDT.......199B})
derives the kinematical parameters from the observed velocity field
by fitting tilted-ring models.
The observed velocities given in the velocity maps, $V_{\mathrm{obs}}$,
are obtained by solving the following equation
$$
V_{\mathrm{obs}} = V_{\mathrm{sys}} + V_{\mathrm{rot}}(R)\cos\theta\sin i + V_{\mathrm{exp}}(R)\sin\theta\sin i
$$
where $V_{\mathrm{rot}}$ is the rotation velocity, $V_{\mathrm{exp}}$
the expansion velocity, $R$ and $\theta$ the polar coordinates in the plane
of the galaxy and $i$ the inclination. The same procedure was used for all
the galaxies in the sample. The physical width of the rings
is always the same: 4.83$''$ for the OmM, 2.04$''$
for the OHP and 1.44$''$ for the CFHT data; in order
to have a good sampling of the signal.

 Since all the galaxies in the \BB sample are barred,
it was decided to derive the kinematical parameters only in the axisymmetric part
of the disc of the galaxy. This means that the central regions of the galaxy
were systematically masked to avoid contamination from non circular
motions due to the presence of the bar. Obviously, the solution will still be affected by the non-circular motions of the spiral disc itself. In order to determine the range of
galactic radii to apply \textsc{ROTCUR}, the values of the deprojected bar lengths given
by \citeauthor{1995AJ....109.2428M} (\citeyear{1995AJ....109.2428M}, and reported in Table \ref{tablesample}, column 9) were used.
For five galaxies not belonging to the \ggo Martin's\ggf\ sample (IC 0342, NGC 1530,
NGC 2403, NGC 4535 and NGC 5964), an ellipse fitting was done to determine
an approximate bar length.

For each galaxy in the sample, a fit is first done simultaneously for
the kinematical centre $(x_{pos},y_{pos})$ and the systemic
velocity $V_{sys}$, fixing the position angle $P.A.$ and the
inclination $i$ (using the photometrical values).

Secondly, by keeping $V_{sys}$ and the kinematical
centre fixed, $P.A.$ and $i$ are allowed to
vary over the same radius range and their mean values derived.
It was decided to use the mean values of $P.A.$ and $i$ since discs are rarely warped inside
the optical radius; warps are mainly seen for R $>$ R$_{opt}$.
$V_\mathrm{rot}$ are then calculated keeping the 5 derived kinematical parameters,
$\left (x_{pos},y_{pos}\right )$,
$V_{sys}$, $P.A.$ and $i$ fixed over the whole radius range. In all
the fits, $V_{\mathrm{exp}}$ was not considered and fixed to zero. Finally,
a 2D kinematical model for each galaxy of the \BB sample was constructed
using the \textsc{velfi} routine of \textsc{GIPSY} and subtracted
from the data to get a residual velocity map.
This whole process was repeated until the mean and the dispersion of residuals were found to be minimal and close to zero and with a distribution as homogeneous as possible over the whole FOV.

%For all the galaxies, it is the case, except in the central regions were
%non-circular motions due to the bar are found.
To quantify the effect of masking the bar in the determination of the disc parameters, the residual maps (including the bar and masking it) were systematically compared. The means and the dispersions were computed for each residual map and for each galaxy. Results are presented in Table \ref{AR}. The difference between the mean of the residual map in the disc part, using the model with the bar, and the mean of the residual map in the disc part, using the model avoiding the bar, are always close to zero (except for NGC7479 where there is a severe perturbation in the westerm arm). A difference is noticeable when comparing the dispersion of the residuals in the disc part, using the model with the bar and the dispersion of the residuals in the disc part, using the model avoiding the bar. The latter is always smaller than the former ($<$20\kms). This suggests that the determination of the disc parameters is better when the bar region is avoided.
\begin{center}
\begin{table*}
\caption{Analysis of the residuals}
\label{AR}
\begin{tabular}{c|llll}
\hline \hline
%   Galaxy & m$_{in disc with bar}$ - m$_\mathrm{in disc avoiding bar}$  & Dispersion of residuals $_\mathrm{in disc with bar}$& Dispersion of residu$_\mathrm{in disc avoiding bar}$ \\ 
   Galaxy & $\overline{m}_{idwb}$ - $\overline{m}_{idab}$  & $\overline{\sigma}_{idwb}$& $\overline{\sigma}_{idab}$ &  Remarks \\ 
      & (\kms) & (\kms) & (\kms) & \\
\hline
  NCG 0925  &    +5.2 & 30.6 & 17.0 \\
 \phantom{NI}IC  0342  &  n/a & n/a & n/a  & bar not well defined\\
  NGC 1530  &    +11.4 & 51.6& 31.6&  \\
  NGC 2336  &  n/a & n/a & n/a  & no \ha in the bar region   \\
  NGC 2403  &   n/a & n/a & n/a  & bar not well defined \\
  NCG 2903  &  +1.0 & 20.3 & 19.8   \\
  NGC 3198  &  +0.20 & 12.0 & 10.6 &  \\
  NCG 3359  &  +1.0 & 21.5 & 12.9 \\
  NGC 3953  &  +2.2 & 27.3 & 14.0 \\
  NGC 3992  &   n/a & n/a & n/a  & no \ha in the bar region   \\
  NGC 4236  &   5.8 & 21.4 & 10.8 \\
  NGC 4321  &    +0.8 & 15.6 & 13.8 \\
  NGC 4535  &    +0.7 & 16.5 & 13.4 \\
  NGC 5371 &  n/a & n/a & n/a & no \ha in the bar region \\
  NGC 5457  &  1.1 & 33.7 & 33.0  \\
  NGC 5921 &   n/a & n/a & n/a  & no \ha in the bar region   \\
  NGC 5964  &  +2.8 & 25.9 & 16.7 &   \\
  NGC 6217  &  +2.6 & 20.7 & 13.8  \\
  NGC 6946  &  +1.1 & 18.5 & 13.3 \\
  NGC 7479  &  +30.2 & 108.7 & 103.56 & "broken arm" in the disc region \\
  NGC 7741  &  +4.9 & 19.5 & 14.5  \\
\hline
\end{tabular}
\\
\begin{flushleft}
$\overline{m}_{idwb}$ : mean of the residual map in the disc part, using the model with the bar\\
$\overline{m}_{idab}$ : mean of the residual map in the disc part, using the model avoiding the bar\\
$\overline{\sigma}_{idwb}$ : dispersion of the residual map in the disc part, using the model with the bar \\
$\overline{\sigma}_{idab}$ : dispersion of the residual map in the disc part, using the model avoiding the bar \\
\end{flushleft}
\end{table*}
\end{center}

Once the most suitable 2D kinematical model was found, the
\textsc{kpvslice} routine of \textsc{karma} was used to derive a
Position-Velocity (PV) diagram. This PV diagram is useful to check if
the rotation curve derived from the whole 2-D velocity field is a
good representation of the kinematics on the major axis. As the axisymmetrical models were derived using the disc region avoiding the bar, it is clear that strong non axisymmetrical motions can be seen in the bar region (e.g. NGC0925, NGC2903, NGC3198, NGC3359, NGC6217 and NGC7741). When the
range of velocities in a galaxy was superior to the Free Spectral
Range of the etalon used, the overall velocity range was 
distributed over two or three orders. In this case, 
data cubes have been replicated in the spectral dimension (twice or three times) to
construct the PV diagrams. This is the case for NGC 7479, NGC 5371,
NGC 4535, NGC 3992, NGC 3953, NGC 2903, NGC 2336 and NGC 1530. 

Table \ref{resultats} gives the results for the kinematical parameters 
fitting, compares with the photometrical data of the RC3 and indicates the
shift between the photometrical and kinematical centres. Since, in the \textsc{ROTCUR} task, 
the kinematical value of $P.A.$ is
defined as the angle measured counterclockwise 
from the North to the receding side of the
velocity field, it may thus differs
from the RC3 value by 180\Deg. 

\begin{figure*}
\begin{center}
\includegraphics[width=7.5cm]{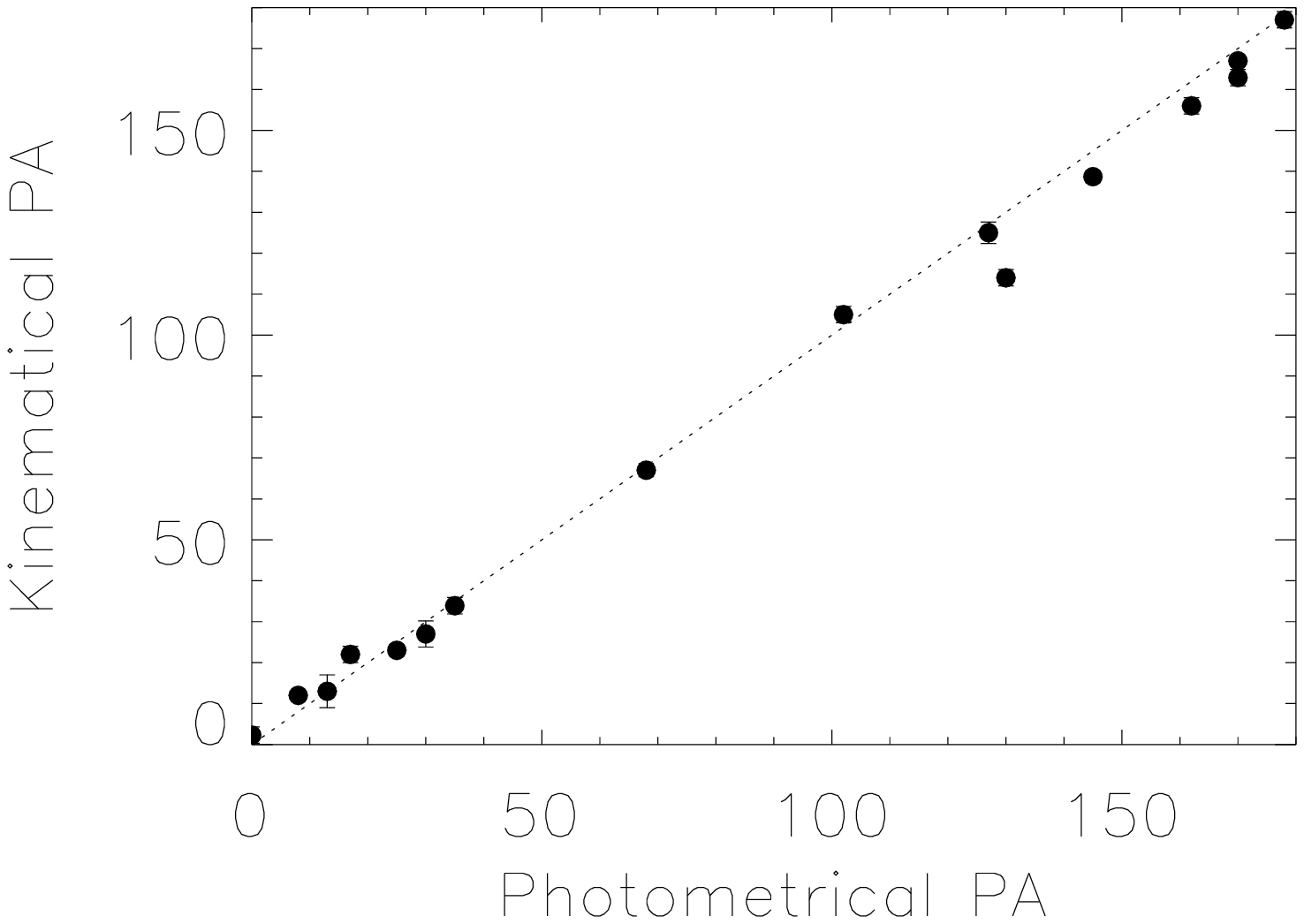} \includegraphics[width=7.5cm]{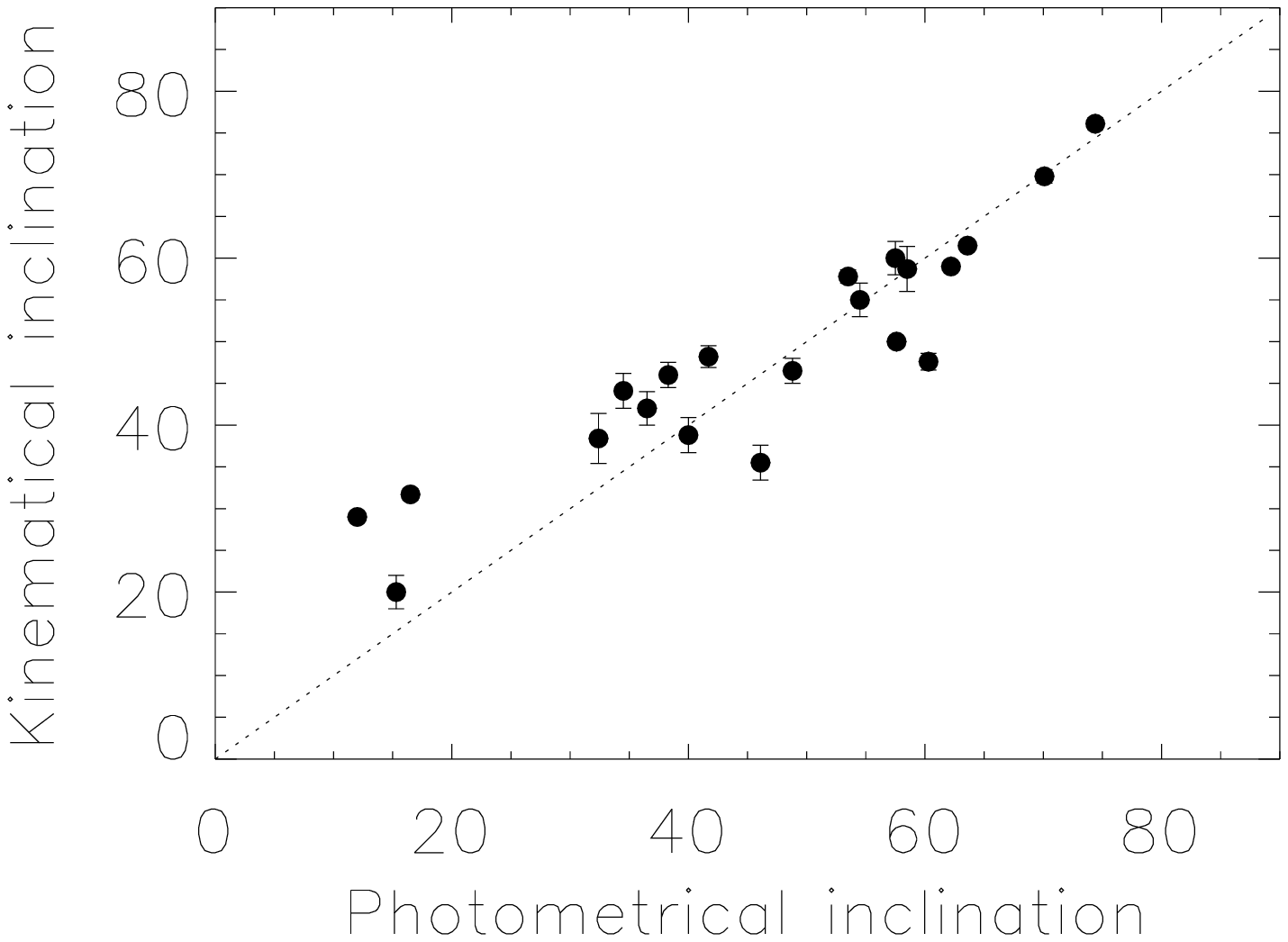}
\caption[Comparison between the kinematical and photometrical position angles and
inclinations]{Comparison between the kinematical and photometrical position angles and
inclinations. Position angles are given modulo 180\Deg.  The dashed line
represents the y=x equation.} \label{pakinpaphot}
\end{center}
\end{figure*}

The agreement is good between the photometrical and the kinematical
position angle, even for low inclination galaxies for which the
photometrical position angle is difficult to determine accurately (Fig \ref{pakinpaphot} left).

\begin{figure*}
\begin{center}
\includegraphics[width=7cm]{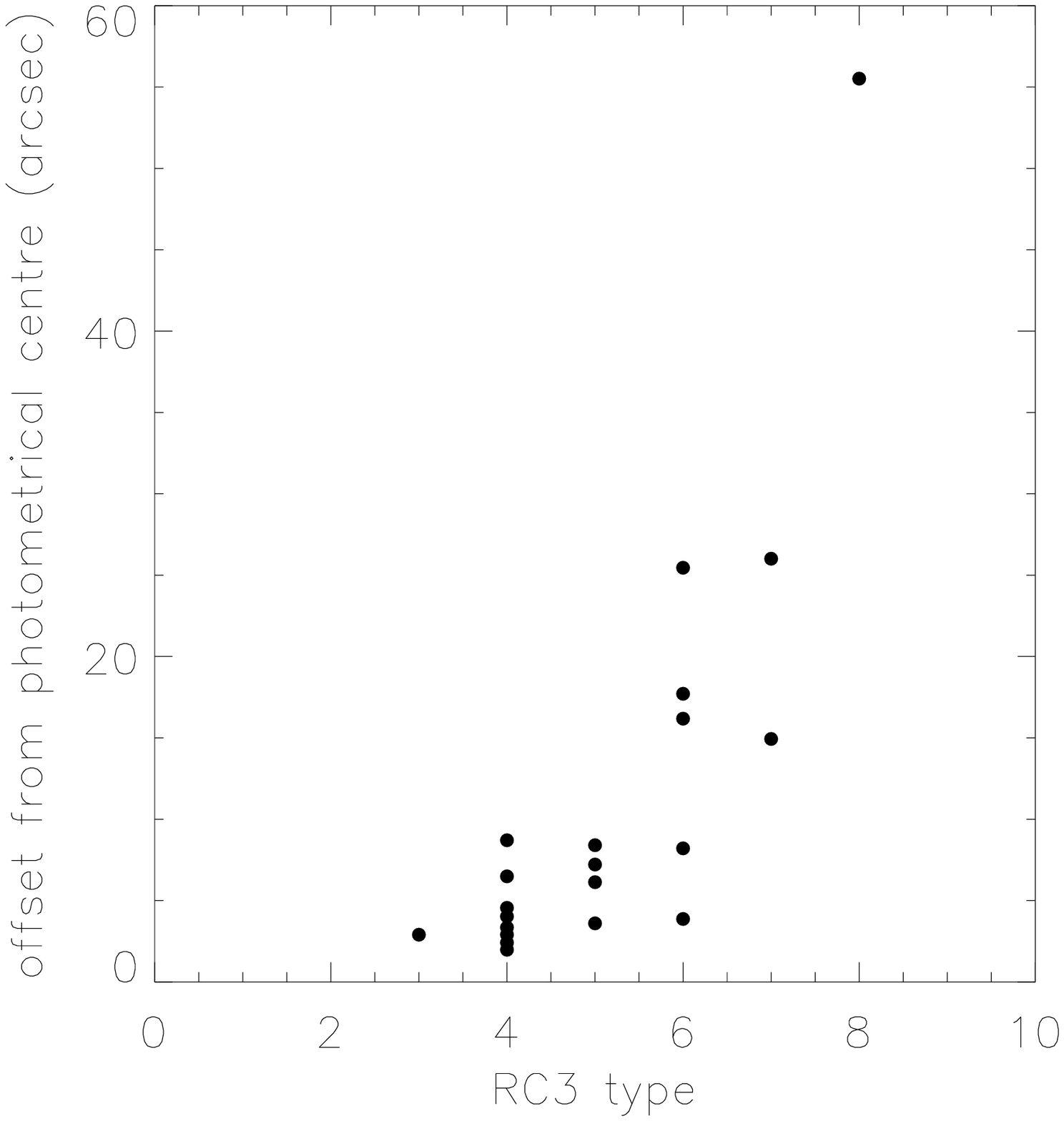} \includegraphics[width=7cm]{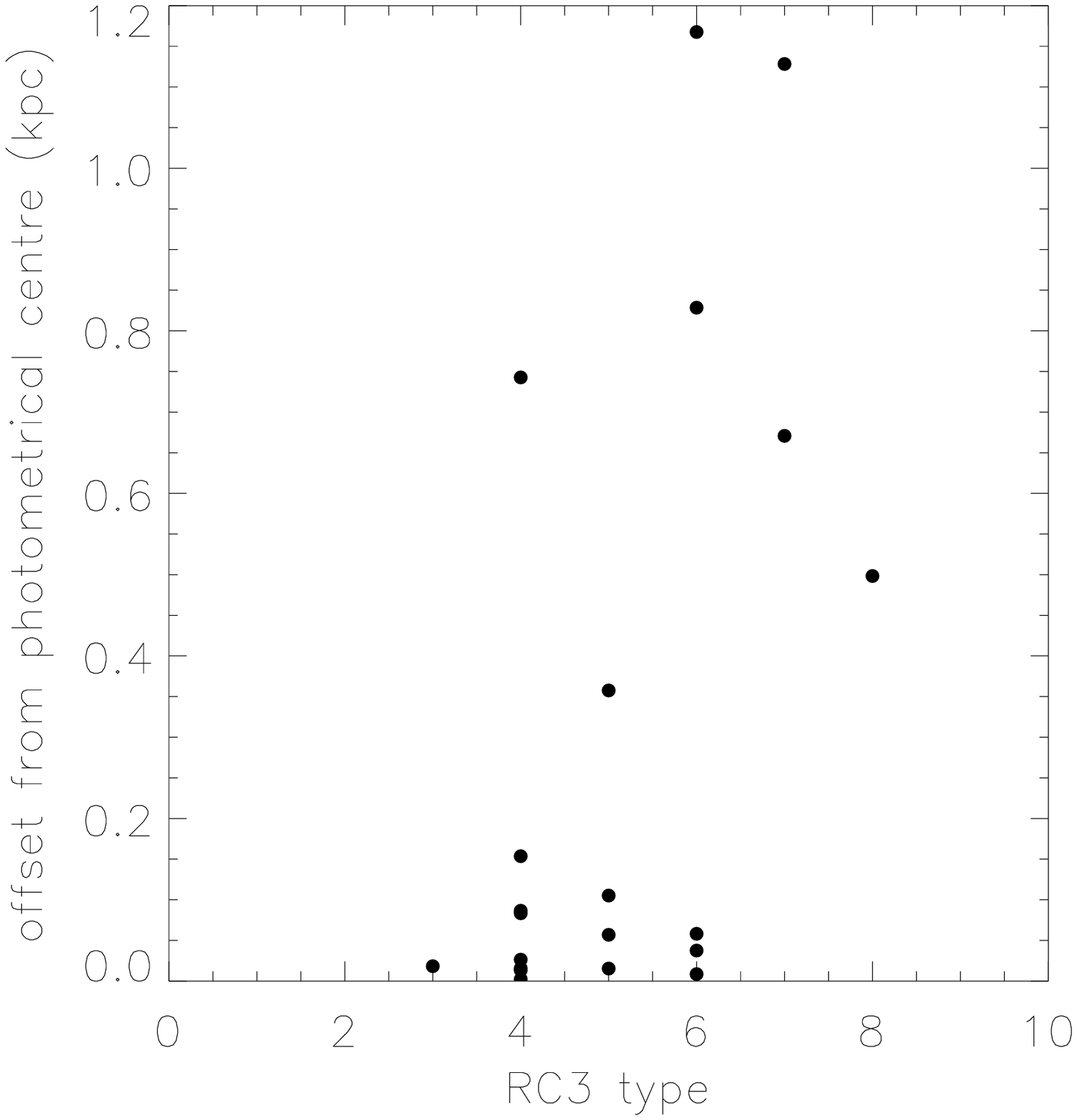}
\caption[Distance between the photometrical and kinematical 
centres as a function of morphological type]{\begin{footnotesize}\textbf{Left}: Distance (in arcsec) between the photometrical and kinematical 
centres as a function of morphological type.  The offsets are
computed in the plane of the sky (not corrected for inclination).
\textbf{Right}: Deprojected distance (in kpc) between the photometrical and kinematical 
centres as function of the morphological type. The offsets have been
computed in the plane of the galaxy, using the position angles and
the inclinations given in Table \ref{resultats}.\end{footnotesize}} \label{center}
\end{center}
\end{figure*}

When comparing inclinations, noticeable discrepancies can be seen for two galaxies having a
round shape (IC 0342 and NGC 4321) and a low photometrical
inclination ($<$20\Deg). The photometrical method used to determine the
inclination, by fitting ellipses to the outer isophotes (or simply from the axis ratios), minimizes the
inclination while the kinematical method is much less sensitive to
this \ggo face-on\ggf\ effect. The shift between the position of the
centre of the galaxy determined from the photometry and the kinematics
is clearly a function of the morphological type of the galaxy (see Fig.~\ref{center}).
The strongest discrepancies occur for later type spirals for which
the photometrical centre is not always easy to identify. The large
value of the offset when plotted in units of \emph{arcsec} on Fig.~\ref{center} (left panel) 
shows that, for a large majority of galaxies, this difference may
not be explained by seeing or spatial resolution effects. On the
other hand, this offset is not statistically significant: NGC 2403,
IC 0342, NGC 0925, and NGC 5457 have an offset of 1.17, 0.89, 0.67 and 0.30~kpc respectively
while the bulk of 15 galaxies over 19 has an offset lower or equal to 0.1 kpc (Note : only 19 galaxies over 21 were studied because
the determination of the photometrical centre for NGC 4236 and NGC 5964 was too hazardous). \\

Future N-body models coupled to a SPH code will
help understand the gas behavior in a non axysimmetrical potential
and may provide an explanation for such differences. The full analysis will be presented later.

\begin{table*}
\caption{HI, kinematical and photometrical position angles and inclinations.
Offsets between kinematical and photometrical centres.} \label{resultats}
\begin{center}
  \begin{tabular}{rrrrrrrrrrr}
  \hline\hline
   Galaxy  &  \multicolumn{2}{c}{HI} & Ref. & \multicolumn{2}{c}{Photometrical} & \multicolumn{2}{c}{\ha kinematical} &\multicolumn{2}{c} {\small Offset from photo. centre} \\
   name &   $P.A.$ (\Deg) & Incl.(\Deg) & & $P.A.$ (\Deg) & Incl.(\Deg) & $P.A.$ (\Deg)& Incl.(\Deg) & (arcsec) & (kpc)  \\
 \hline
  NGC 0925 & 101.0$\pm$1.0* & 58.5 & 1 &  102 & 57.6 & 105.0$\pm$1.0* &  57.8$\pm$1.5 & 26.0 & 0.670\\   %pa = 75
 IC  0342 & 37$\pm$1.0 & 31$\pm$6.0 & 2 &  n/a& 12.0 & 42.0$\pm$2.0 &  29.0$\pm$0.4 & 16.2 &  0.830 \\
 NGC 1530 & n/a& 58 & 3 &n/a & 60.3 & 5.0$\pm$1.0* &  47.6$\pm$0.6 & 2.9 & 0.020\\
 NGC 2336 & n/a & 59 &4& 178& 58.5 & 177.0$\pm$1.3$\dagger$ &  58.7$\pm$2.7 & 4.1 & 0.002 \\
 NGC 2403 & 124& 61& 5 &127& 57.5 & 125.0$\pm$1.0 & 60.0$\pm$2.0 & 25.4 & 1.167\\
 NGC 2903 & 20* & 61& 4  &17& 63.6 & 22.0$\pm$1.0* &  61.5$\pm$0.5 & 2.4 & 0.083 \\ %pa = 202
 NGC 3198 & 36*& 71 &7 & 35& 70.1 & 33.9$\pm$0.3* &  69.8$\pm$0.8 & 8.4 & 0.105  \\
 NGC 3359 & 172$^{\dagger}$ & 51& 8 & 170& 54.5 & 167.0$\pm$2.0$\dagger$ &  55.0$\pm$2.0 & 6.1 & 0.015\\  %pa = 193
 NGC 3953 & 11 & 58 & 9 &13& 62.2 & 13.0$\pm$0.8 & 59.0$\pm$0.4 & 6.5 &  0.086\\ %pa =167
 NGC 3992 & 70$\pm$5*& 57$\pm$1& 10 & 68& 53.5 & 67.0$\pm$1.0* & 57.8$\pm$0.8 & 2.9 & 0.015\\ %pa = 293
 NGC 4236 & 163& 75 & 11 & 162& 74.4 & 156.1$\pm$1.6 & 76.1$\pm$0.7 & 55.5 & 0.498 \\ %pa = 23.9
 NGC 4321 & 27$\pm$1$^{\dagger}$ & 27 & 12 & 30& 16.5 & 27.0$\pm$1.0$\dagger$ &  31.7$\pm$0.7 & 8.7 & 0.101 \\ %pa =153
 NGC 4535 & 3*& 40& 13 & 0& 46.1 & 2.3$\pm$0.4* &  35.0$\pm$2.0 & 3.6 & 0.057   \\
 NGC 5371 & 8 & 43 & 14 & 8& 38.3 & 12.0$\pm$1.0 & 46.0$\pm$1.5 & 2.0 & 0.011\\
 NGC 5457 & 49& 18& 15 & n/a& 15.3& 53.0$\pm$1.0 &20.0$\pm$2.0 & 17.7 & 0.297 \\
 NGC 5921 & n/a & 24  & 16 & 130& 36.5&  114.0$\pm$0.7 &42.0$\pm$2.0 & 4.5 & 0.045\\
 NGC 5964 & 144 & 41 & 17 &145& 40.0& 138.7$\pm$0.7 & 38.8$\pm$2.1 & 15.0 & 1.130 \\
 NGC 6217 & 249& 29& 18 & n/a& 34.5& 250.0$\pm$1.0&  34.1$\pm$2.1 & 3.34 & 0.014\\
 NGC 6946 & 240$\pm$1& 38$\pm$5& 19 & n/a& 32.4& 239.0$\pm$1.0& 38.4$\pm$3.0  &  8.2 & 0.163\\
 NGC 7479 & 22& 51& 20 & 25& 41.7&  23.0$\pm$1.0* & 48.2$\pm$1.3 & 7.2 & 0.060\\ 
 NGC 7741 & 47 & 5 & 4 & 170& 48.8& 162.9$\pm$0.5* & 46.5$\pm$1.5 & 3.9 & 0.023\\
\hline 
\end{tabular}
\begin{flushleft}
%\end{flushleft}
$P.A.$ = Position Angle of the major axis of the galaxy. $\dagger$ indicates that $P.A.$ = 180\Deg - kinematical $P.A.$. 
$*$ indicates that $P.A.$=180 + kinematical$P.A.$\\
 References for HI data : 1 - Pisano, Wilcots \& Elmegreen (1998); 2 - Crosthwaite, Turner \& Ho. (2000); 3 - \citeauthor{1993AAS...183.7606T} (\citeyear{1993AAS...183.7606T}); 4 - WHISP data (\citeauthor{2001ASPC..240..451V} \citeyear{2001ASPC..240..451V}); 5 - \citeauthor{2000A&A...356L..49S} (\citeyear{2000A&A...356L..49S}); 8- \citeauthor{1986ApJ...307..453B}  (\citeyear{1986ApJ...307..453B}); 9 - \citeauthor{2001A&A...370..765V} (\citeyear{2001A&A...370..765V} ); 10 - \citeauthor{2002A&A...388..793B} (\citeyear{2002A&A...388..793B}); 11 - \citeauthor{1973A&A....24..411S} (\citeyear{1973A&A....24..411S}); 12 - \citeauthor{1993ApJ...416..563K}  (\citeyear{1993ApJ...416..563K}); 13- \citeauthor{1988AJ.....96..851G} (\citeyear{1988AJ.....96..851G}); 14 - \citeauthor{1987SvAL...13..186Z} (\citeyear{1987SvAL...13..186Z}); 15 - \citeauthor{1981A&A....93..106B}  (\citeyear{1981A&A....93..106B}); 16 - \citeauthor{1994ApJ...423..180S} (\citeyear{1994ApJ...423..180S}); 17 - \citeauthor{1983AJ.....88..272H} (\citeyear{1983AJ.....88..272H}); 18 - \citeauthor{1991A&A...245....7V} (\citeyear{1991A&A...245....7V}); 19 - \citeauthor{1990A&A...234...43C} (\citeyear{1990A&A...234...43C}); 20 - \citeauthor{1998MNRAS.297.1041L} (\citeyear{1998MNRAS.297.1041L}) 
\end{flushleft}
\end{center}
\end{table*}

\section{Concluding remarks and further work.}

The 3D data presented in this paper are the results of a survey of
the \ha kinematics of nearby barred galaxies with the FP
integral-field spectrometer \FM. This study provides an homogenous
sample of barred galaxies.  The 3D data were processed through a
robust reduction pipeline.  An adaptive binning method has been used to achieve
optimal spatial coverage and resolution at a given signal-to-noise ratio, typically
around 7. High spatial and spectral resolutions \ha monochromatic
maps and velocity fields are presented.  Bars' signatures in
velocity fields and position-velocity diagrams, reveal strong non
circular motions and thus provides observational constrains to
extract the parameters of the bar and of the disc. Fine tuning of
position angles and inclinations have been done.
The kinematical parameters have been determined using a tilted-ring
model, by taking into account only the axisymmetric part of the disc
to avoid any kind of contamination due to non-circular motions
from the bar (in the inner parts of the disc) or to a possible warp (in the outer parts).

The analysis of the sample shows that the photometrical and
kinematical parameters (position angle of the major axis,
inclination and centre) are in relative good agreement, except
maybe for the later-type spirals.  Nevertheless, the determination
of the kinematical parameters is more accurate than the photometrical
one. 

The main purpose of the paper is to provide and present an
homogenous 3D data sample of nearby barred galaxies useful for
further analysis. Velocity patterns of the bar(s) and of the
spiral will be accurately determined using the Tremaine-Weinberg 
method (Tremaine \& Weinberg 1984) on the H$\alpha$ velocity fields in two forthcoming papers.
In Paper II the Tremaine-Weinberg method applied to the gaseous
component will be discussed using numerical simulations and
illustrated with the galaxy M100. In Paper III, the rest of the
sample will be analyzed using this Tremaine-Weinberg method.
In paper IV, the rotation curves will be derived, properly corrected for
non-circular motions in order to retrieve the actual mass
distributions.  This will be achieved with N-body coupled with SPH
simulations for each velocity fields of the \BB sample. This kind
of approach has been recently used by \citeauthor{2004A&A...424..799P}
(\citeyear{2004A&A...424..799P}). However, our study will differ
in two major aspects. First, their comparison is relative to a
small sample, whereas \BB is homogeneous and well distributed over
the Hubble sequence. Second, they only used long-slit data,
whereas \ha velocity fields and \ha monochromatic images will be
used to provide more accurate results avoiding, for example, the
uncertainties on $P.A.$ and $i$ present in 1D data. This will be the
first step toward the determination of accurate mass models for
barred spiral galaxies, until the time when full 2D mass models
using the whole velocity fields will become available.

The whole data presented in the paper will be available for the community once paper II to IV will be published.

\section{Acknowledgments}

We thank Chantal Balkowski for her help and support at the different
stages of this work. We also thank Olivia Garrido, Jacques
Boulesteix, Jean-Luc Gach, Philippe Balard and Olivier Boissin for
their support. It is a pleasure to thank the OmM staff, in
particular Bernard Malenfant and Ghislain Turcotte for their
enthusiastic and competent support. Let us, also, acknowledge
Pierre Martin and the CFHT team for their support at the top of
the Mauna Kea. The \FM\, project has been carried out by the
Laboratoire d'Astrophysique Exp\'erimentale
(\textbf{\texttt{LAE}}) of the Universit\'e de Montr\'eal using a
grant from the Canadian Foundation for Innovation and the
Minist\`ere de l'Education du Qu\'ebec. This project made use of
the LEDA database: http://leda.univ-lyon1.fr/. The Digitized Sky
Surveys were produced at the Space Telescope Science Institute
under U.S. Government grant NAG W-2166. The images of these
surveys are based on photographic data obtained using the Oschin
Schmidt Telescope on Palomar Mountain and the UK Schmidt
Telescope.

\begin{figure*}
  \begin{center}
    \vspace{9cm} \begin{Large}\textbf{High resolution images avaliable on}\end{Large} \\ 
     \vspace{1cm} \begin{Large}\texttt{www.astro.umontreal.ca/fantomm/bhabar/}\end{Large} \vspace{7cm}
  \end{center}
\caption[NGC 0925]{NGC 0925. \textbf{Top left} : XDSS Blue Band image. \textbf{Top right} : 3.6 $\mu$m Spitzer image. \textbf{Middle left} : \ha monochromatic image. \textbf{Middle right} : \ha velocity field. \textbf{Bottom} : PV diagram. The \ha velocity field and \ha monochromatic image were spatially binned using an adaptive binning. The scale of the color used to represent the velocity is located to the right of the \ha map. The PV diagram was obtained integrating over a
slit of 3 pixels wide using the velocities of the model extracted from GIPSY and the data cube from the observations.}
\label{n0925}
\end{figure*}
\clearpage

\begin{figure*}
  \begin{center}
\vspace{9cm} \begin{Large}\textbf{High resolution images avaliable on}\end{Large} \\ 
     \vspace{1cm} \begin{Large}\texttt{www.astro.umontreal.ca/fantomm/bhabar/}\end{Large} \vspace{7cm}\
  \end{center}
\caption[IC 0342]{\begin{small}IC 0342. \textbf{Top left} : XDSS Blue Band image. \textbf{Top right} : 2Mass Ks-band image. \textbf{Middle left} : \ha monochromatic image. \textbf{Middle right} : \ha velocity field. \textbf{Bottom} : PV diagram.  \end{small}}
\label{n0342}
\end{figure*}
\clearpage

\begin{figure*}
  \begin{center}
\vspace{9cm} \begin{Large}\textbf{High resolution images avaliable on}\end{Large} \\ 
     \vspace{1cm} \begin{Large}\texttt{www.astro.umontreal.ca/fantomm/bhabar/}\end{Large} \vspace{7cm}\

  \end{center}
\caption[NGC 1530]{NGC 1530. \textbf{Top left} : XDSS Blue Band image. \textbf{Top right} : 2Mass Ks-band image. \textbf{Middle left} : \ha monochromatic image. \textbf{Middle right} : \ha velocity field. \textbf{Bottom} : PV diagram.}
\label{n1530}
\end{figure*}
\clearpage

\begin{figure*}
  \begin{center}
    \vspace{9cm} \begin{Large}\textbf{High resolution images avaliable on}\end{Large} \\ 
     \vspace{1cm} \begin{Large}\texttt{www.astro.umontreal.ca/fantomm/bhabar/}\end{Large} \vspace{7cm}\
  \end{center}
\caption[NGC 2336]{NGC 2336. \textbf{Top left} : XDSS Blue Band image. \textbf{Top right} : 2Mass Ks-band image. \textbf{Middle left} : \ha monochromatic image. \textbf{Middle right} : \ha velocity field. \textbf{Bottom} : PV diagram.}
\label{n2336}
\end{figure*}
\clearpage

\begin{figure*}
  \begin{center}
\vspace{9cm} \begin{Large}\textbf{High resolution images avaliable on}\end{Large} \\ 
     \vspace{1cm} \begin{Large}\texttt{www.astro.umontreal.ca/fantomm/bhabar/}\end{Large} \vspace{7cm}\
  \end{center}
\caption[NGC 2403]{NGC 2403. \textbf{Top left} : XDSS Blue Band image. \textbf{Top right} : 3.6 $\mu$m Spitzer image. \textbf{Middle left} : \ha monochromatic image. \textbf{Middle right} : \ha velocity field. \textbf{Bottom} : PV diagram.}
\label{n2403}
\end{figure*}
\clearpage

\begin{figure*}  
   \begin{center}
\vspace{9cm} \begin{Large}\textbf{High resolution images avaliable on}\end{Large} \\ 
     \vspace{1cm} \begin{Large}\texttt{www.astro.umontreal.ca/fantomm/bhabar/}\end{Large} \vspace{7cm}\
  \end{center}
\caption[NGC 2903]{NGC 2903. \textbf{Top left} : XDSS Blue Band image. \textbf{Top right} : 2Mass Ks band image. \textbf{Middle left} : \ha monochromatic image. \textbf{Middle right} : \ha velocity field. \textbf{Bottom} : PV diagram.}
\label{n2903}
  \end{figure*}
\clearpage

\begin{figure*}  
 \begin{center}
\vspace{9cm} \begin{Large}\textbf{High resolution images avaliable on}\end{Large} \\ 
     \vspace{1cm} \begin{Large}\texttt{www.astro.umontreal.ca/fantomm/bhabar/}\end{Large} \vspace{7cm}\
  \end{center}
\caption[NGC 3198]{NGC 3198. \textbf{Top left} : XDSS Blue Band image. \textbf{Top right} : 3.6 $\mu$m Spitzer image. \textbf{Middle left} : \ha monochromatic image. \textbf{Middle right} : \ha velocity field. \textbf{Bottom} : PV diagram.}
\label{n3198}
  \end{figure*}
\clearpage

\begin{figure*}  
   \begin{center}
\vspace{9cm} \begin{Large}\textbf{High resolution images avaliable on}\end{Large} \\ 
     \vspace{1cm} \begin{Large}\texttt{www.astro.umontreal.ca/fantomm/bhabar/}\end{Large} \vspace{7cm}\
  \end{center}
\caption[NGC 3359]{NGC 3359. \textbf{Top left} : XDSS Blue Band image. \textbf{Top right} : 2Mass Ks band image. \textbf{Middle left} : \ha monochromatic image. \textbf{Middle right} : \ha velocity field. \textbf{Bottom} : PV diagram.}
\label{n3359}
  \end{figure*}
\clearpage

\begin{figure*}  
  \begin{center}
\vspace{9cm} \begin{Large}\textbf{High resolution images avaliable on}\end{Large} \\ 
     \vspace{1cm} \begin{Large}\texttt{www.astro.umontreal.ca/fantomm/bhabar/}\end{Large} \vspace{7cm}\
  \end{center}
\caption[NGC 3953]{NGC 3953. \textbf{Top left} : XDSS Blue Band image. \textbf{Top right} : 2Mass Ks band image. \textbf{Middle left} : \ha monochromatic image. \textbf{Middle right} : \ha velocity field. \textbf{Bottom} : PV diagram.}
\label{n3953}
  \end{figure*}
\clearpage

\begin{figure*}  
  \begin{center}
\vspace{9cm} \begin{Large}\textbf{High resolution images avaliable on}\end{Large} \\ 
     \vspace{1cm} \begin{Large}\texttt{www.astro.umontreal.ca/fantomm/bhabar/}\end{Large} \vspace{7cm}\
  \end{center}
\caption[NGC 3992]{NGC 3992. \textbf{Top left} : XDSS Blue Band image. \textbf{Top right} : 2Mass Ks band image. \textbf{Middle left} : \ha monochromatic image. \textbf{Middle right} : \ha velocity field. \textbf{Bottom} : PV diagram.}
\label{n3992}
  \end{figure*}
\clearpage

\begin{figure*}  
  \begin{center}
\vspace{9cm} \begin{Large}\textbf{High resolution images avaliable on}\end{Large} \\ 
     \vspace{1cm} \begin{Large}\texttt{www.astro.umontreal.ca/fantomm/bhabar/}\end{Large} \vspace{7cm}\
  \end{center}
\caption[NGC 4236]{NGC 4236. \textbf{Top left} : XDSS Blue Band image. \textbf{Top right} : 3.6 $\mu$m Spitzer image. \textbf{Middle left} : \ha monochromatic image. \textbf{Middle right} : \ha velocity field. \textbf{Bottom} : PV diagram.}
\label{n4236}
  \end{figure*}
\clearpage

\begin{figure*}  
 \begin{center}
\vspace{9cm} \begin{Large}\textbf{High resolution images avaliable on}\end{Large} \\ 
     \vspace{1cm} \begin{Large}\texttt{www.astro.umontreal.ca/fantomm/bhabar/}\end{Large} \vspace{7cm}\
  \end{center}
\caption[NGC 4321]{NGC 4321. \textbf{Top left} : B band image from Knapen et al. 2004. \textbf{Top right} : 3.6 $\mu$m Spitzer image. \textbf{Middle left} : \ha monochromatic image. \textbf{Middle right} : \ha velocity field. \textbf{Bottom} : PV diagram.}
\label{n4321}
  \end{figure*}
\clearpage

\begin{figure*}  
  \begin{center}
\vspace{9cm} \begin{Large}\textbf{High resolution images avaliable on}\end{Large} \\ 
     \vspace{1cm} \begin{Large}\texttt{www.astro.umontreal.ca/fantomm/bhabar/}\end{Large} \vspace{7cm}\
  \end{center}
\caption[NGC 4535]{NGC 4535. \textbf{Top left} : B band image from Knapen et al. 2004. \textbf{Top right} : Ks band image from Knapen et al. 2003. \textbf{Middle left} : \ha monochromatic image. \textbf{Middle right} : \ha velocity field. \textbf{Bottom} : PV diagram.}
\label{n4535}
  \end{figure*}
\clearpage

\begin{figure*}  
  \begin{center}
\vspace{9cm} \begin{Large}\textbf{High resolution images avaliable on}\end{Large} \\ 
     \vspace{1cm} \begin{Large}\texttt{www.astro.umontreal.ca/fantomm/bhabar/}\end{Large} \vspace{7cm}\
  \end{center}
\caption[NGC 5371]{NGC 5371. \textbf{Top left} : B band image from Knapen et al. 2004. \textbf{Top right} : Ks band image from Knapen et al. 2003. \textbf{Middle left} : \ha monochromatic image. \textbf{Middle right} : \ha velocity field. \textbf{Bottom} : PV diagram.}
\label{n5371}
  \end{figure*}
\clearpage

\begin{figure*}  
 \begin{center}
\vspace{9cm} \begin{Large}\textbf{High resolution images avaliable on}\end{Large} \\ 
     \vspace{1cm} \begin{Large}\texttt{www.astro.umontreal.ca/fantomm/bhabar/}\end{Large} \vspace{7cm}\
  \end{center}
\caption[NGC 5457]{NGC 5457. \textbf{Top left} : XDSS Blue Band image. \textbf{Top right} : 2Mass Ks band image. \textbf{Middle left} : \ha monochromatic image. \textbf{Middle right} : \ha velocity field. \textbf{Bottom} : PV diagram.}
\label{n5457}
  \end{figure*}
\clearpage

\begin{figure*}  
 \begin{center}
\vspace{9cm} \begin{Large}\textbf{High resolution images avaliable on}\end{Large} \\ 
     \vspace{1cm} \begin{Large}\texttt{www.astro.umontreal.ca/fantomm/bhabar/}\end{Large} \vspace{7cm}\
  \end{center}
\caption[NGC 5921]{NGC 5921. \textbf{Top left} : XDSS Blue Band image. \textbf{Top right} : 2Mass Ks band image. \textbf{Middle left} : \ha monochromatic image. \textbf{Middle right} : \ha velocity field. \textbf{Bottom} : PV diagram.}
\label{n5921}
  \end{figure*}
\clearpage

\begin{figure*}  
  \begin{center}
\vspace{9cm} \begin{Large}\textbf{High resolution images avaliable on}\end{Large} \\ 
     \vspace{1cm} \begin{Large}\texttt{www.astro.umontreal.ca/fantomm/bhabar/}\end{Large} \vspace{7cm}\
  \end{center}
\caption[NGC 5964]{NGC 5964. \textbf{Top left} : B band image from Knapen et al. 2004. \textbf{Top right} : Ks band image from Knapen et al. 2003. \textbf{Middle left} : \ha monochromatic image. \textbf{Middle right} : \ha velocity field. \textbf{Bottom} : PV diagram.}
\label{n5964}
  \end{figure*}
\clearpage

\begin{figure*}  
 \begin{center}
\vspace{9cm} \begin{Large}\textbf{High resolution images avaliable on}\end{Large} \\ 
     \vspace{1cm} \begin{Large}\texttt{www.astro.umontreal.ca/fantomm/bhabar/}\end{Large} \vspace{7cm}\
  \end{center}
\caption[NGC 6217]{NGC 6217. \textbf{Top left} : XDSS Blue Band image. \textbf{Top right} : 2Mass Ks band image. \textbf{Middle left} : \ha monochromatic image. \textbf{Middle right} : \ha velocity field. \textbf{Bottom} : PV diagram.}
\label{n6217}
  \end{figure*}
\clearpage

\begin{figure*}  
 \begin{center}
\vspace{9cm} \begin{Large}\textbf{High resolution images avaliable on}\end{Large} \\ 
     \vspace{1cm} \begin{Large}\texttt{www.astro.umontreal.ca/fantomm/bhabar/}\end{Large} \vspace{7cm}\
  \end{center}
\caption[NGC 6946]{NGC 6946. \textbf{Top left} : B band image from Knapen et al. 2004. \textbf{Top right} : 3.6 $\mu$m Spitzer image. \textbf{Top right} : 2Mass Ks band image. \textbf{Middle left} : \ha monochromatic image. \textbf{Middle right} : \ha velocity field. \textbf{Bottom} : PV diagram.}
\label{n6946}
  \end{figure*}
\clearpage

\begin{figure*}  
  \begin{center}
\vspace{9cm} \begin{Large}\textbf{High resolution images avaliable on}\end{Large} \\ 
     \vspace{1cm} \begin{Large}\texttt{www.astro.umontreal.ca/fantomm/bhabar/}\end{Large} \vspace{7cm}\
  \end{center}
\caption[NGC 7479]{NGC 7479. \textbf{Top left} : XDSS Blue Band image. \textbf{Top right} : 2Mass Ks band image. \textbf{Middle left} : \ha monochromatic image. \textbf{Middle right} : \ha velocity field. \textbf{Bottom} : PV diagram.}
\label{n7479}
  \end{figure*}
\clearpage

\begin{figure*}  
 \begin{center}
\vspace{9cm} \begin{Large}\textbf{High resolution images avaliable on}\end{Large} \\ 
     \vspace{1cm} \begin{Large}\texttt{www.astro.umontreal.ca/fantomm/bhabar/}\end{Large} \vspace{7cm}\
  \end{center}
\caption[NGC 7741]{NGC 7741. \textbf{Top left} : B band image from Knapen et al. 2004. \textbf{Top right} : Ks band image from Knapen et al. 2003. \textbf{Middle left} : \ha monochromatic image. \textbf{Middle right} : \ha velocity field. \textbf{Bottom} : PV diagram.}
\label{n7741}
  \end{figure*}

\appendix

\section{Description of the individual galaxies}

A brief description of the structures observed in the \ha velocity fields and 
monochromatic images of the \BB sample is given in this section.\\

\textbf{IC  0342:} This large SAB(rs)cd nearby galaxy is
nearly face-on so its photometrical position angle ($P.A.$) is
uncertain. A strong difference between its kinematical and
photometrical inclination is noted. Observations in CO and \hi\ 
(\citeauthor{2000AJ....119.1720C} \citeyear{2000AJ....119.1720C}) suggest that its kinematical $P.A.$ is
37\Deg. The kinematical data from CO, \hi\ and \ha are comparable.
An \ha spiral structure near the centre can be seen.

\textbf{NGC 0925:} This late type SBcd galaxy has a bright optical
and \ha bar and two bright patchy spiral arms beginning at the
ends of the bar. Many \hii\ regions lie along the bar. Photometrical
and kinematical data agree. The PV diagram shows non axisymmetric
motions near the centre. It is well studied in \hi\ 
(\citeauthor{1998ApJ...494L..37E} \citeyear{1998ApJ...494L..37E},
\citeauthor{1998AJ....115..975P} \citeyear{1998AJ....115..975P}),
in
 CO (\citeauthor{2003ApJS..145..259H} \citeyear{2003ApJS..145..259H}) and
in \ha (\citeauthor{1982A&A...108..134M} \citeyear{1982A&A...108..134M}).
It shows strong streaming motions.

\textbf{NGC 1530:} This strongly barred spiral galaxy has been extensively studied by
Zurita et al. 2004. Here the \FM\, observation provides much more details of the
kinematics along the major axis due to its large FOV and high sensivity. A nuclear spiral and a large velocity gradient are observed.
In the V-band images, a nuclear bar (\citeauthor{1993AJ....105.1344B} \citeyear{1993AJ....105.1344B}) with hot spots (\citeauthor{1973PASP...85..103S} \citeyear{1973PASP...85..103S}) can be
seen. 

\textbf{NGC 2336:} NGC 2336 is an intermediate-type ringed barred spiral galaxy with a prominent bar. 
This galaxy shows a very regular morphological structure with no major asymmetries, except for the central regions where HI maps show a lack
of gas in the centre (\citeauthor{1983A&AS...54...19V}
\citeyear{1983A&AS...54...19V}). The \ha monochromatic image suggests the same lack. The NW part
of the disc has been cutoff by the wings of the interference filter. Although NGC 2336
belongs to an apparent pair of
galaxies (together with IC 467) with a projected linear distance of 135 kpc, its undisturbed
disc does not exhibit any distinct sign of recent interactions
(\citeauthor{1999A&A...344..787W} \citeyear{1999A&A...344..787W}).

\textbf{NGC 2403:} This SABc galaxy shows amorphous spiral features. The \ha velocity maps
and the PV diagram show an almost rigid structure near the centre of the galaxy. Bright \hii\ 
regions can be seen in the \ha monochromatic image. It is not clear whether this galaxy in barred
or not. According to  \citeauthor{1997MNRAS.292..349S} (\citeyear{1997MNRAS.292..349S}), their
Fourier harmonic analysis of the \hi\ velocity field shows that non-circular motions are not
important in this galaxy.
Moreover, \citeauthor{2000A&A...356L..49S} (\citeyear{2000A&A...356L..49S}) stress that
the thin hydrogen disc of NGC 2403 is surrounded by a vertically extended layer of \hi\ that
rotates slower than the disc. A complete modeling of the galaxy will provide more details
on its structures. \citeauthor{2001ApJ...562L..47F}  (\citeyear{2001ApJ...562L..47F}) suggest
that this anomalous \hi\ component may be similar to a class of high velocity clouds observed
in the Milky Way. In CO data, no molecular gas is detected (Helfer et al. 2003).

\textbf{NGC 2903:} This starburst galaxy shows several peaks of star formation in the
circumnuclear region. These peculiar \ggo hotspots\ggf\ have been identified
and described in different ways by various authors in \ha (\citeauthor{1983A&A...128..140M}
\citeyear{1983A&A...128..140M}), in radio and in the infrared by \citeauthor{1985ApJ...290..108W}
(\citeyear{1985ApJ...290..108W}). A strong velocity gradient can be seen along the bar in the
velocity map. The PV diagram shows a clear step in the RC. This step-like structure could be
related to its strong bar. The molecular gas, visible in CO observations (Helfer et al. 2003),
follows the bar.

\textbf{NGC 3198:}
This SB(rs)c galaxy has been extensively studied in \hi\ (Bosma 1981a, \citeauthor{1989A&A...223...47B} \citeyear{1989A&A...223...47B}),
FP \ha (\citeauthor{1991A&A...244...27C} \citeyear{1991A&A...244...27C},
Blais-Ouellette et al. 1999) and \ha and [\nii] long-slit spectroscopy
(\citeauthor{1998PASJ...50..427S} \citeyear{1998PASJ...50..427S}, \citeauthor{2004AJ....127.3273V} \citeyear{2004AJ....127.3273V}).
According to the PV diagram, non circular motions  near the centre can be seen. A strong
velocity gradient is also seen perpendicular to the bar major axis.

\textbf{NGC 3359:}
NGC 3359 is a strongly barred galaxy. Its \hii\ regions have been studied
(\citeauthor{1995ApJ...445..161M} \citeyear{1995ApJ...445..161M}) and cataloged
(\citeauthor{2000A&A...354..823R} \citeyear{2000A&A...354..823R}). The structure and
kinematics of the \hi\ were analyzed in detail by \citeauthor{1982AJ.....87..751G}
(\citeyear{1982AJ.....87..751G}) and \citeauthor{1986ApJ...307..453B}
(\citeyear{1986ApJ...307..453B}) showing a clumpy distribution and a 
low surface density within the annular zone of strong star formation, which can be
explained as due to the effect of the bar sweeping up gas as it rotates. Analysis of
the \ha velocity map shows a disc with an axisymmetric rotation and also evidence of
strong non circular motions as confirmed by \citeauthor{2000A&AS..142..259R}
(\citeyear{2000A&AS..142..259R}). Analysis of the \ha residual velocity map (not shown here) 
shows strong streaming motions in the spiral arms and a strong gradient of gas in the bar.

\textbf{NGC 3953:} One of the most massive spirals of the M81 group, it has been studied in \hi\ 
by \citeauthor{2001A&A...370..765V} (\citeyear{2001A&A...370..765V}).
It is rather poor in \hi\ and the surface density drops near $D_{25}$. Moreover, in CO (Helfer et al. 2003),
the gas seems to be located in a ring at the near end of the bar. In \ha, the same lack
of gas in the centre can be seen, whereas the arms are well developped.

\textbf{NGC 3992:} This galaxy is the most massive spiral of the M81 group.
Its \hi\ distribution is regular. It has a prominent bar and very well defined
spiral arms. It has a faint radial \hi\ extension outside its stellar disc. There is a
pronounced central \hi\ hole in the gas distribution at exactly the radial extent of the
bar (\citeauthor{2002A&A...388..793B} \citeyear{2002A&A...388..793B}), also visible in
the \ha emission line. Some \ha can be seen toward the centre of the galaxy.
It is not clear whether this feature could exist in \hi\ because of the poor spatial resolution
of the \hi\ data measurements. Observations in CO (Helfer et al. 2003) stress the lack of
molecular gas in the galaxy.

\textbf{NGC 4236:} This late type SBdm galaxy is seen nearly edge-on. Its kinematical
inclination is 76\Deg. The \ha image shows that the \hii\ regions are distributed along the bar,
with two bright regions near the end of the bar. These features are also seen in
\hi\ (\citeauthor{1973A&A....24..411S} \citeyear{1973A&A....24..411S}). An extensive region
of solid-body rotation coincides with the bar.

\textbf{NGC 4321:} This grand-design spiral galaxy has been frequently mapped in the $H\alpha$
emission line using high-resolution FP
interferometry (\citeauthor{1990A&A...234...23A} \citeyear{1990A&A...234...23A},
\citeauthor{1990A&AS...83..211C} \citeyear{1990A&AS...83..211C}, \citeauthor{1997ApJ...479..723C} \citeyear{1997ApJ...479..723C},
\citeauthor{2000ApJ...528..219K}  \citeyear{2000ApJ...528..219K}), in the molecular CO
emission-line (\citeauthor{1990PhDT.........6C} \citeyear{1990PhDT.........6C},
\citeauthor{1995AJ....110.2075S} \citeyear{1995AJ....110.2075S}, \citeauthor{1995AJ....109.2444R}
\citeyear{1995AJ....109.2444R}, \citeauthor{1998A&A...333..864G} \citeyear{1998A&A...333..864G},
Helfer et al. 2003) and in the 21-cm \hi\ emission-line (\citeauthor{1990AJ....100..604C}
\citeyear{1990AJ....100..604C}, \citeauthor{1993ApJ...416..563K} \citeyear{1993ApJ...416..563K}).
The \hi\ disc is almost totally confined within the optical one but with a slight lopsideness
towards the SW (\citeauthor{1993ApJ...416..563K} \citeyear{1993ApJ...416..563K}).
The \hi, CO and \ha velocity fields show
kinematical disturbances such as streaming motions along the spiral arms and a central S-shape
distortion of the iso-velocity contours along the bar axis. The circum-nuclear region (CNR)
and shows the presence of an enhanced star formation region as a four-armed \ha ring-like structure
and a CO \& \ha spiral-like structure. Much more details can be found in Hernandez et al. 2005.

\textbf{NGC 4535:} With NGC 4321, this is another Virgo cluster galaxy with a known Cepheids
distance (16.0 Mpc). It is not located in the main sub-structure of the
cluster close to the core elliptical galaxy M87 but it lies in the southern
extension related to M49. Although the \ha morphology appears perturbed,
with an obvious indication of multiple spiral arms (these features are even
present in the NIR Ks image), its velocity field is regular. No \hi\ was
found in the central parts (Cayatte et al. 1990).

\textbf{NGC 5371:} In this galaxy, there is no evidence of \ha emission in the centre.

\textbf{NGC 5457:} M101 is a large nearby galaxy. Observations of CO data (BIMA song, Helfer
et al. 2003) show that the molecular gas is only distributed along the bar. In the \ha image the gas is distributed over the whole field. Two large arms can be seen.

\textbf{NGC 5921:} This galaxy shows a ring like structure in both the \ha and
the XDSS blue images. The bar is not clearly seen in \ha but many \hii\ regions lie at the end of the bar. The centre of
galaxy is the host of a strong velocity gradient well defined in the \ha velocity map. Nevertheless, no \ha gas is present between the ring like structure and the centre.

\textbf{NGC 5964:} This galaxy shows strong \hii\ regions distributed along its bar. Nevertheless, no
strong velocity gradient can be seen in this direction.

\textbf{NGC 6217:} This galaxy presents a prominent bar, with an important region of stellar
formation on it located at 10\Arc\ in the southeastern direction from the galactic centre
(\citeauthor{1999ARep...43..377A} \citeyear{1999ARep...43..377A}). The galactic centre reveals
the presence of a different structural component, a ring. The size of the ring in \ha  
observations is approximately of 43\Arc, which agrees with previous works in \hi\ on this galaxy (\citeauthor{1991A&A...245....7V} \citeyear{1991A&A...245....7V}). The inner ring structure in the centre of the galaxy can be easily seen in the \ha monochromatic image.

\textbf{NGC 6946:}
According to \hi\ studies (\citeauthor{1990A&A...234...43C} \citeyear{1990A&A...234...43C}), the
\hi\ distribution is not symmetric but is more extended to the NE side. This feature is also
seen in the \ha  emission map. The overall \ha velocity map is regular but shows some
non-circular motions near the centre, confirmed by the PV diagram. It has been recently observed
in FP by (\citeauthor{blaisous2004} \citeyear{blaisous2004}) leading to the same
conclusions. Once again the wide field of \FM\ and its high sensitivity is clearly an advantage
to obtain better \ha velocity fields.

\textbf{NGC 7479:}
 This SB(s)c galaxy has been studied in \hi\ by \citeauthor{1998MNRAS.297.1041L}
(\citeyear{1998MNRAS.297.1041L}). The \hi\ distribution shows considerable asymmetries
and distortions in the outer disc. The \hi\ and \ha kinematics suggest that, while the
global velocity field is fairly regular, a severe perturbation is present in the western
spiral arm.
There is also a strong velocity gradient along the bar confirmed in the PV diagram.
The \ha monochromatic image shows strong \hii\ regions at one end of the bar.

\textbf{NGC 7741:} This galaxy shows a strong bar in the \has, blue and Ks band
images. Many \hii\ regions are located along the bar. The velocity map clearly shows
non-circular motions, whereas the velocity gradient is not too strong (compared with NGC 2903
for example).

\bsp

\label{lastpage}

\end{document}